\documentclass{jfm}
\usepackage{graphicx}
\usepackage{amsmath}
\usepackage{overpic}
\usepackage{float}
\usepackage{epstopdf,epsfig}
\usepackage{color,contour}
\usepackage{amssymb}
\usepackage[dvipsnames]{xcolor}
\usepackage{graphicx}
\usepackage{tabularx}
\usepackage[dvipsnames]{xcolor}
\usepackage{booktabs}
\usepackage[colorlinks,citecolor = blue, linkcolor=red,hyperindex,CJKbookmarks]{hyperref}

\shorttitle{Penetrative turbulent Rayleigh-B\'enard convection}
\shortauthor{Q. Wang, Q. Zhou, Z. Wan, and D. Sun}

\title{Penetrative turbulent Rayleigh-B\'enard convection in two and three dimensions}

\author{Qi Wang\aff{1},
 Quan Zhou\aff{2},
  Zhen-Hua Wan\aff{1}$\dag$
and
 De-Jun Sun\aff{1}
\corresp{\email{wanzh@ustc.edu.cn;dsun@ustc.edu.cn}}
}
\affiliation{\aff{1}Department of Modern Mechanics, University of Science and Technology of China,
Hefei, 230027, China
\aff{2}Shanghai Institute of Applied Mathematics and Mechanics, Shanghai University, Shanghai 200072, China}

\begin{document}

\maketitle

\begin{abstract}
Penetrative turbulent Rayleigh-B\'enard convection which depends on the density maximum of water near $4^\circ\rm{C}$ is studied using two-dimensional (2D) and three-dimensional (3D) direct numerical simulations (DNS). The working fluid is water near $4^\circ\rm{C}$ with Prandtl number $Pr=11.57$.
The considered Rayleigh numbers $Ra$ range from $10^7$ to $10^{10}$. The density inversion parameter $\theta_m$ varies from 0 to 0.9.
It is found that the ratio of the top and bottom thermal boundary-layer thickness ($F_\lambda=\lambda_t^\theta/\lambda_b^\theta$) increases with increasing $\theta_m$, and the relationship between $F_\lambda$ and $\theta_m$ seems to be independent of $Ra$. The centre temperature $\theta_c$ is enhanced compared to that of Oberbeck-Boussinesq (OB) cases, as $\theta_c$ is related to $F_\lambda$ with $1/\theta_c=1/F_\lambda+1$,  $\theta_c$ is also found to have a universal relationship with $\theta_m$ which is independent of $Ra$. Both the Nusselt number $Nu$ and the Reynolds number $Re$ decrease with increasing $\theta_m$, the normalized Nusselt number $Nu(\theta_m)/Nu(0)$ and Reynolds number $Re(\theta_m)/Re(0)$ also have universal relationships with $\theta_m$ which seem to be independent of both $Ra$ and the aspect ratio $\Gamma$. The scaling exponents of $Nu\sim Ra^\alpha$ and $Re\sim Ra^\beta$ are found to be insensitive to $\theta_m$ despite of the remarkable change of the flow organizations.
\end{abstract}


\section{Introduction}
Penetrative convection refers to the phenomena whenever convection in a thermally unstable fluid layer penetrates into adjacent stable layers. The unstable fluid layers are often bounded by rigid boundaries in most laboratory experiments on convection. However, stellar convection zones are bounded by stably-stratified regions and the understanding of penetration of convection across the interface between stable and unstable layers is of astrophysical importance. For instance,  cold plumes from the outer convective zone of the sun penetrate into the upper layers of the tachocline and generate internal gravity waves, this process is thought to play an important role in the turbulent transport of momentum in the tachocline \citep{dintrans2005spectrum}. The observable motions in the outer stable regions of the sun presumably arise from the solar convective zone which is just below them \citep{leighton1963solar}. The origin of Jupiter's zonal winds remains a puzzle and there are two different views: the shallow convection model and the deep convection model \citep{kong2018origin}. The deep convection model thinks  that the zonal flow originates from the penetrative convection that takes place in the deep hydrogen-helium interior \citep{zhang1996penetrative,zhang2000teleconvection}. Although Earth's liquid core is convectively unstable to convection, some studies show that the outermost part of the Earth's core may be  stably-stratified \citep{buffett2014geomagnetic}. The extent of the penetration of convection into the stable layers is important since anomalous depletion of lithium in late stars might be explained if convection penetrates deep enough for the Li-He reaction to occur \citep{antar1987penetrative}.

Penetrative convection is usually studied in simplified model problems. \cite{goluskin2016penetrative} considered an internally heated fluid layer which is confined between top and bottom plates of equal temperatures. The unstably-stratified upper region drives convection that penetrates into the stably-stratified lower region. Another commonly used model to study penetrative convection is convection of water near $4^\circ\rm{C}$. Consider a Rayleigh-B\'enard  convection (RBC)\citep{ahlers2009heat,lohse2010small,chilla2012new} system using water as the working fluid, the temperature of the bottom plate is higher than $4^\circ\rm{C}$ while the temperature of the top plate is lower than $4^\circ\rm{C}$. The fluid in the lower layer above the $4^\circ\rm{C}$ level is convectively unstable while the upper layer below the $4^\circ\rm{C}$ level is gravitationally stable, any convection in the lower, unstable region will penetrate into the stable layer above, and this will lead to the so-called penetrative convection.  Convection of water near $4^\circ\rm{C}$ was often used as a model problem to study penetrative convection in the past.

The first theoretical work on the subject of penetrative convection considered a layer of water in which the bottom boundary is maintained at $0^\circ\rm{C}$ and the top boundary is kept at a temperature larger than $4^\circ\rm{C}$ \citep{veronis1963penetrative}. A criteria is built for the onset of instability, and it is demonstrated that the system can become unstable to a finite amplitude disturbance at values of the Rayleigh number $Ra$ less than the critical value of the infinitesimal stability theory.  This nonlinear behaviour was later studied by \cite{musman1968penetrative}. \cite{moore1973nonlinear} used numerical simulations to extend the small-amplitude solution of penetrative convection to higher $Ra$. \cite{large2014penetrative} experimentally investigate penetrative RBC, they found that the conduction-convection transition is hysteretic in nature, and the flow undergoes secondary transitions to either hexagonal cellular or longitudinal roll states at higher $Ra$. \cite{hu2015rayleigh} investigated RBC of water near $4^\circ\rm{C}$ in a cubical cavity with different thermal boundaries on the sidewalls and found that there are multiple flow pattern coexistence and hysteresis phenomenon during the flow pattern transition.

Most of the previous works are conducted at relatively low $Ra$ where the flow is laminar. There are less works devoted to turbulent penetrative convection. \cite{lecoanet2015numerical} studied internal gravity wave excitation by convection of water near $4^\circ\rm{C}$ at $Ra=5.8\times10^7$ using two-dimensional (2D) simulations. A recent work \citep{toppaladoddi2018penetrative} investigated 2D penetrative convection at relatively high Rayleigh numbers $10^6\le Ra\le10^8$ using lattice Boltzmann method, they adopted periodical conditions in the horizontal direction and the Prandtl numbers are 1 and 11.6. More recently, \cite{couston2018order} numerically investigated 2D penetrative convection with horizontal periodical conditions and found that a periodic, oscillating mean flow spontaneously
develops from turbulently generated internal waves.  The non-Oberbeck-Boussinesq (NOB) nature of water near $4^\circ\rm{C}$ will cause a top-down symmetry breaking, leading to a shift of centre temperature $\theta_c$, i.e. $\theta_c$ will deviate from that of the OB case. This deviation has been investigated in RBC with NOB effects caused by large temperature differences \citep{zhang1997non,ahlers2006non,ahlers2007non,sameen2008non,sameen2009specific,sugiyama2009flow,horn2013non,horn2014rotating,weiss2018bulk}. However, there is still a lack of systematic studies on the centre temperature $\theta_c$ in turbulent penetrative RBC of water near $4^\circ\rm{C}$. We try to fill this gap and report universal relationship of $\theta_c$ with the density inversion parameter $\theta_m$. The ratio of the top and bottom thermal boundary-layer (BL) thicknesses $F_\lambda=\lambda_t^\theta/\lambda_b^\theta$, the normalized Nusselt number $Nu(\theta_m)/Nu(0)$ and Reynolds number $Re(\theta_m)/Re(0)$ are also found to have universal relationships with $\theta_m$ which seem to be independent of $Ra$. 

The rest of the paper is organized as follows. We first briefly describe the governing equations and numerical methods in \S\ref{sec2}. Section \ref{sec3} presents and discusses the universal relationships of $F_\lambda$, $\theta_c$, normalised Nusselt number $Nu(\theta_m)/Nu(0)$ and normalised Reynolds number $Re(\theta_m)/Re(0)$ with respect to $\theta_m$. Finally, we summarize our findings in \S\ref{sec4}, and the simulation details are tabulated in the Appendix.

\begin{figure}
  \centering
  \vskip 2mm
  \begin{overpic}[width=0.8\textwidth]{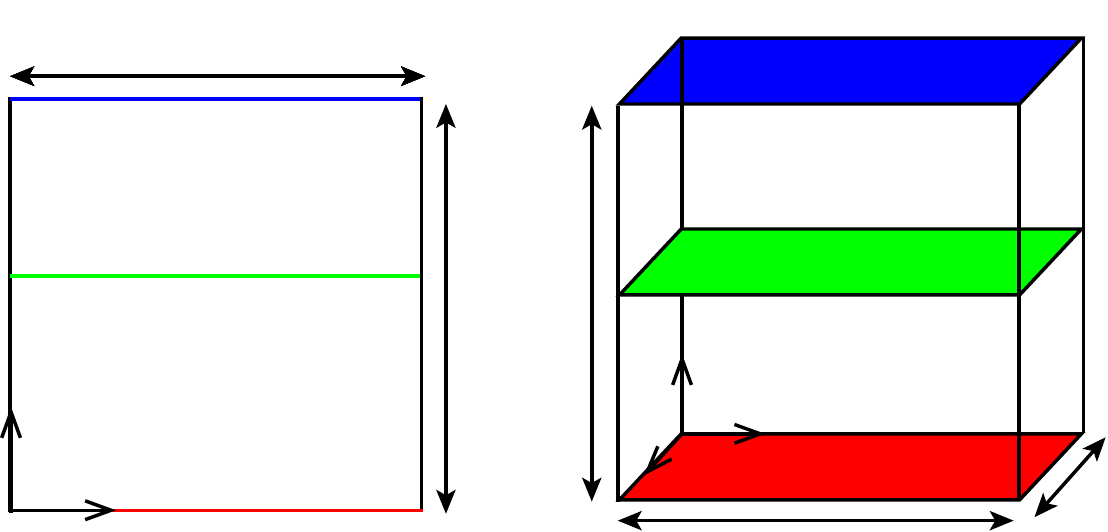}
  \put( -4,40){$(a)$}
  \put(50,40){$(b)$}
  \put( -2,9){$z$}
  \put( 8,-1){$y$}
  \put(15,3){$\hat{T}_b>4^\circ\rm{C}$}
   \put(6,12){Unstable stratification}
  \put(15,20){$\hat{T}_m=4^\circ\rm{C}$}
   \put(6,27){Stable stratification}
  \put(15,35.5){$\hat{T}_t<4^\circ\rm{C}$}
  \put(41,20){$H$}
  \put(19,42){$L$}
  \put(57,7){$x$}
  \put(59,14){$z$}
  \put(66,11){$y$}
   \put(73,-2){$L$}
   \put(50,20){$H$}
    \put(98,2){$W$}
    \end{overpic}
   \vskip 2mm
 \caption{Sketch of (\textit{a}) 2D and (\textit{b}) 3D configurations for penetrative  Rayleigh-B\'enard convection}\label{sketch}
\end{figure}
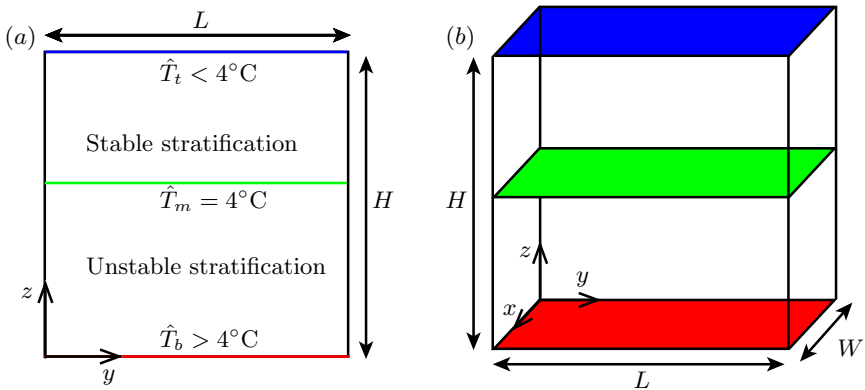

\section{Numerical procedures}\label{sec2}

The problem is sketched in figure \ref{sketch}, the temperature of the bottom plate $\hat{T}_{b}$ (quantities marked with a circumflex are dimensional) is higher than $4^\circ\rm{C}$ while the temperature of the top plate is lower than $4^\circ\rm{C}$. The sidewalls are insulated. We assume thermophysical properties to be constant except the density in the buoyancy term. The nonlinear relationship of density with temperature for cold water near $4^\circ $ is described as
$\hat{\rho}=\hat{\rho}_m(1-\hat{\alpha}{|\hat{T}-\hat{T}_m|}^q)$ \citep{gebhart1977new}, where $\hat{\rho}_m\approx1000kg/m^3$ is the maximum density at the temperature $\hat{T}_m=4^\circ\rm{C}$, the isobaric thermal expansion coefficient $\hat{\alpha}=9.30\times10^{-6}(^\circ {\rm C})^{-q}$, where $q=1.895$. The dimensionless governing equations of this problem read

\begin{gather}
\nabla\cdot\boldsymbol{u} = 0 \label{eq01} \\
\frac{\partial \boldsymbol{u}}{\partial t} + \boldsymbol{u}\cdot\nabla\boldsymbol{u} = -\nabla p+ \sqrt{\frac{Pr}{Ra}}\nabla^2\boldsymbol{u} + |\theta-\theta_m|^q{\vec{\boldsymbol{e}}_z} \label{eq02}\\
\frac{\partial \theta}{\partial t} + \boldsymbol{u}\cdot\nabla \theta  = \frac{1}{\sqrt{RaPr}}\nabla^2\theta\label{eq05}
\end{gather}

\noindent where $\boldsymbol{u}$, $\theta$ and $p$ are the velocity, temperature and  pressure, respectively. For non-dimensionalization, we choose $\hat{H}$ and $\hat{U}={(\hat{g}\hat{\alpha}\Delta\hat{T}^q\hat{H})}^{1/2}$ as reference length and velocity, where $\hat{g}$ is the gravitational acceleration, $\Delta\hat{T}=\hat{T}_{b}-\hat{T}_{t}$ is the temperature difference. The reference time is free-fall time $\hat{t}_{f}=\hat{H}/\hat{U}$. Temperature is nondimensionalized as $\theta=(\hat{T}-\hat{T}_{t})/(\hat{T}_{b}-\hat{T}_{t})$. Three main control parameters are the Prandtl number $Pr$, the Rayleigh number $Ra$, the density inversion parameter $\theta_m$, and they are defined as

\begin{gather}
Pr=\frac{\hat{\nu}}{\hat{\kappa}}, ~~Ra=\frac{\hat{g}\hat{\alpha}(\hat{T}_{b}-\hat{T}_{t})^q\hat{H}^3}{\hat{\nu}\hat{\kappa}}, ~~\theta_m=\frac{\hat{T}_m-\hat{T}_{t}}{\hat{T}_{b}-\hat{T}_{t}}  \label{eq06}
\end{gather}

\noindent where  $\hat{\kappa}$ is the thermal diffusivity, and $\hat{\nu}$ the kinematic viscosity. $\theta_m$ is an important parameter that describes density maximum effect, quantifying the location of $\hat{T}_m$ with respect to wall temperature $\hat{T}_{t}$ and  $\hat{T}_{b}$. For $\theta_m=0$, the top plate is $4^\circ\rm{C}$, and the situation is quite similar to OB approximation. For $\theta_m=1$, the bottom plate is $4^\circ\rm{C}$, and the flow is pure conduction state for any $Ra$. Another control parameter is the aspect ratio. For the 2D configuration, the aspect ration is defined as $\Gamma=\hat{L}/\hat{H}$. For the 3D configuration, the dimensionless length is fixed to 1 and the aspect ratio is defined as $\Gamma_{3}=\hat{W}/\hat{H}$.

The governing equations are solved numerically by an in-house code lMn2d/3d \citep{xia2016flow}. Since our code has been validated and used to study various convection problem before \citep{wang2017thermal,liu2018linear,wang2018multiple,wang2018flow,wang2019flow,wang2019non}, we only give its main features here. All spatial terms are discretised using a second-order central difference scheme.
Non-uniform grids with clustered points near walls are used.
Time integration is accomplished using an Adams-Bashforth scheme for the nonlinear terms and Crank-Nicolson scheme for the viscous and diffusion terms.
Multi-grid strategy is used to solve the pressure Poisson equation.

The $Pr$ is fixed to $11.57$ corresponding to the water at $4^\circ \rm{C}$.  We simulate 2D cases in the range of $10^7\le Ra\le 10^{10}$ and 3D cases for $Ra=10^8$ and $10^9$. Grids which satisfy resolution requirement of DNS \citep{shishkina2010boundary} are used. The details of the main simulations are tabulated in the Appendix. For all the simulations, we generally perform at least 1600 free fall time units. We waited at least 800 free-fall time units before starting to average in order to ensure all transients have been dissipated, and the $Nu$ and $Re$ are averaged for at least 800 free-fall time units. For large $\theta_m$ close to 0.9, longer-time simulations are needed to reach statistically steady state \citep{toppaladoddi2018penetrative}. For example, for the 2D case with $Ra=5\times10^8,\theta_m=0.9, \Gamma=1$, we even preformed 30 000 free-fall time units for the flow to become statistically steady, and another 10 000 free-fall time units are simulated for time average. The results are also compared with OB cases where we set $q=1$ and $\theta_m=0$. Note that for the penetrative case with $\theta_m=0$, the flow is similar to OB cases, but it can not fully recover to OB case since $q=1.895$, and the top-bottom symmetry is still broken. For the OB case, the initial conditions are conduction states superposed with a small perturbation added to the temperature, while for the penetrative cases, we generally use fully-developed flow fields for the OB cases as initial conditions.

\section{Results and discussion}\label{sec3}

\subsection{Flow organisation and centre temperature}

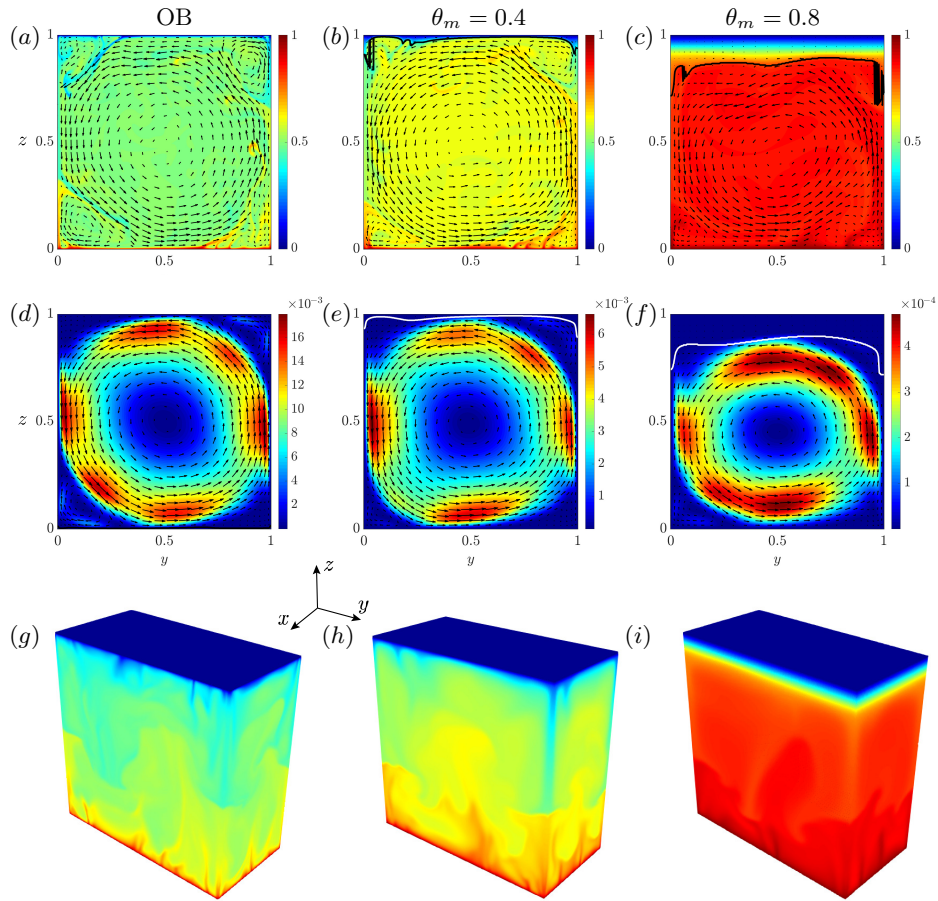
\begin{figure}
  \centering
  \vskip 2mm
  \begin{overpic}[width=0.9\textwidth]{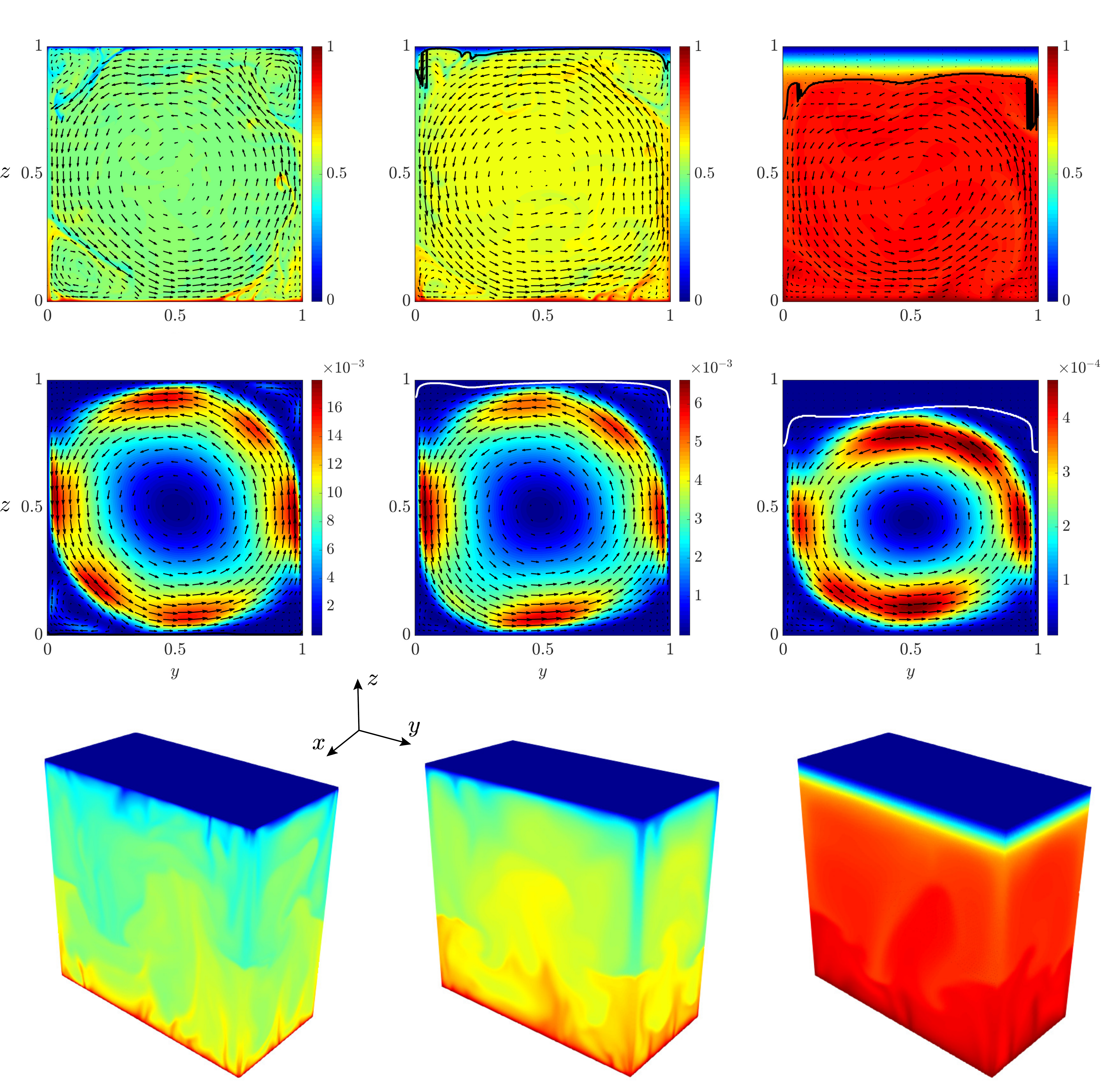}
     \put(-1,94){$(a)$}
     \put(33,94){$(b)$}
     \put(66,94){$(c)$}
     \put(-1,64){$(d)$}
     \put(33,64){$(e)$}
     \put(66,64){$(f)$}
      \put(-1,29){$(g)$}
     \put(33,29){$(h)$}
     \put(66,29){$(i)$}
     \put(15,96){OB}
     \put(45,96){$\theta_m=0.4$}
     \put(77,96){$\theta_m=0.8$}
     \end{overpic}
 \caption{(\textit{a-c}) Instantaneous temperature fields superimposed with velocity vectors for $Ra=10^9$ for 2D cases with $\Gamma=1$. (\textit{d-f}) Time-averaged kinetic energy $1/2(v^2+w^2)$ distribution superimposed with velocity vectors for $Ra=10^9$ for 2D cases with $\Gamma=1$.  (\textit{g-h})  Instantaneous temperature fields for $Ra=10^9$ for 3D cases with $\Gamma_3=1/2$. (\textit{a},\textit{d},\textit{g})  OB. (\textit{b},\textit{e},\textit{h}) $\theta_m=0.4$. (\textit{c},\textit{f},\textit{i}) $\theta_m$=0.8. The black lines in (\textit{b}, \textit{c}) and white lines in (\textit{e}, \textit{f}) denote the location where the  temperature is $4^\circ$.}\label{flow}
\end{figure}

Figure \ref{flow} shows instantaneous temperature (top row)/time-averaged kinetic energy $0.5(v^2+w^2)$ (middle row) fields superimposed with velocity vectors for 2D cases with $Ra=10^9,\Gamma=1$, and instantaneous temperature fields for 3D cases with $Ra=10^9, \Gamma_3=1/2$ (bottom row).  It is seen that a large-scale circulation (LSC) which spans the whole size of the convection cell exists for the OB cases as shown in figures \ref{flow}($a$), \ref{flow}($d$) and \ref{flow}($g$). For the penetrative cases, due to the penetration of the lower hotter unstable fluid layer into the upper stably-stratified layer, the temperature in the bulk increases  compared to that of the OB cases. For relatively small $\textcolor{red}{\theta_m}$, the fluid in the lower layer can penetrate into the whole part of the upper layer as shown in figures \ref{flow}($b$), \ref{flow}($e$) and \ref{flow}($h$). However, for large $\textcolor{red}{\theta_m}$, vigorous convection can not happen in the whole  cell, and the flow is stably-stratified near the top plate as depicted in figures \ref{flow}($c$),  \ref{flow}($f$) and \ref{flow}($i$). The convection is very weak near the top plate, and heat is transferred mainly by thermal conduction in this region. This stably-stratified temperature field is somewhat similar to the stably-stratified angular velocity field in counter-rotating Taylor-Couette flow where Rayleigh-stable zones exist near the outer cylinder \citep{van2011torque,ostilla2013optimal,ostilla2014exploring}. The black lines in figures \ref{flow}($b$) and \ref{flow}($c$) denote the location where the temperature is $4^\circ\rm{C}$. Some fluctuations are observed due  to  the interaction between lower convection region and upper stably-stratified region. The white lines in figures \ref{flow}($e$) and \ref{flow}($f$) also depict the location at which the temperature is $4^\circ \rm{C}$, but in  time-averaged sense. It is clearly shown that the fluctuations disappear after time average. In figures \ref{flow}($d-f$), one can also observe that the velocity magnitude decreases with increasing $\theta_m$.

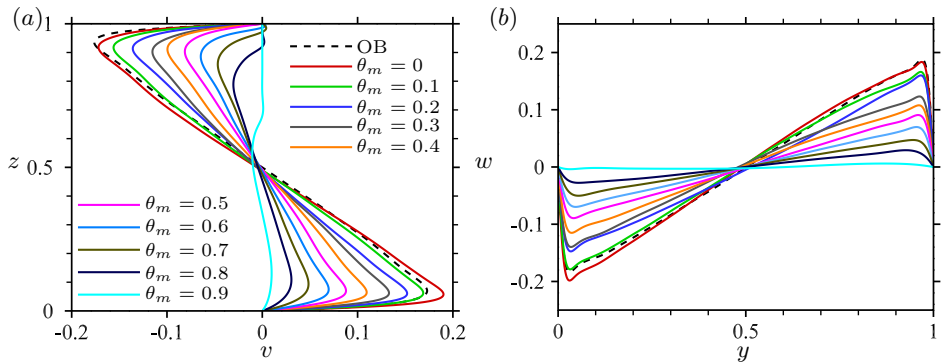
\begin{figure}
  \centering
  \begin{overpic}[width=0.45\textwidth]{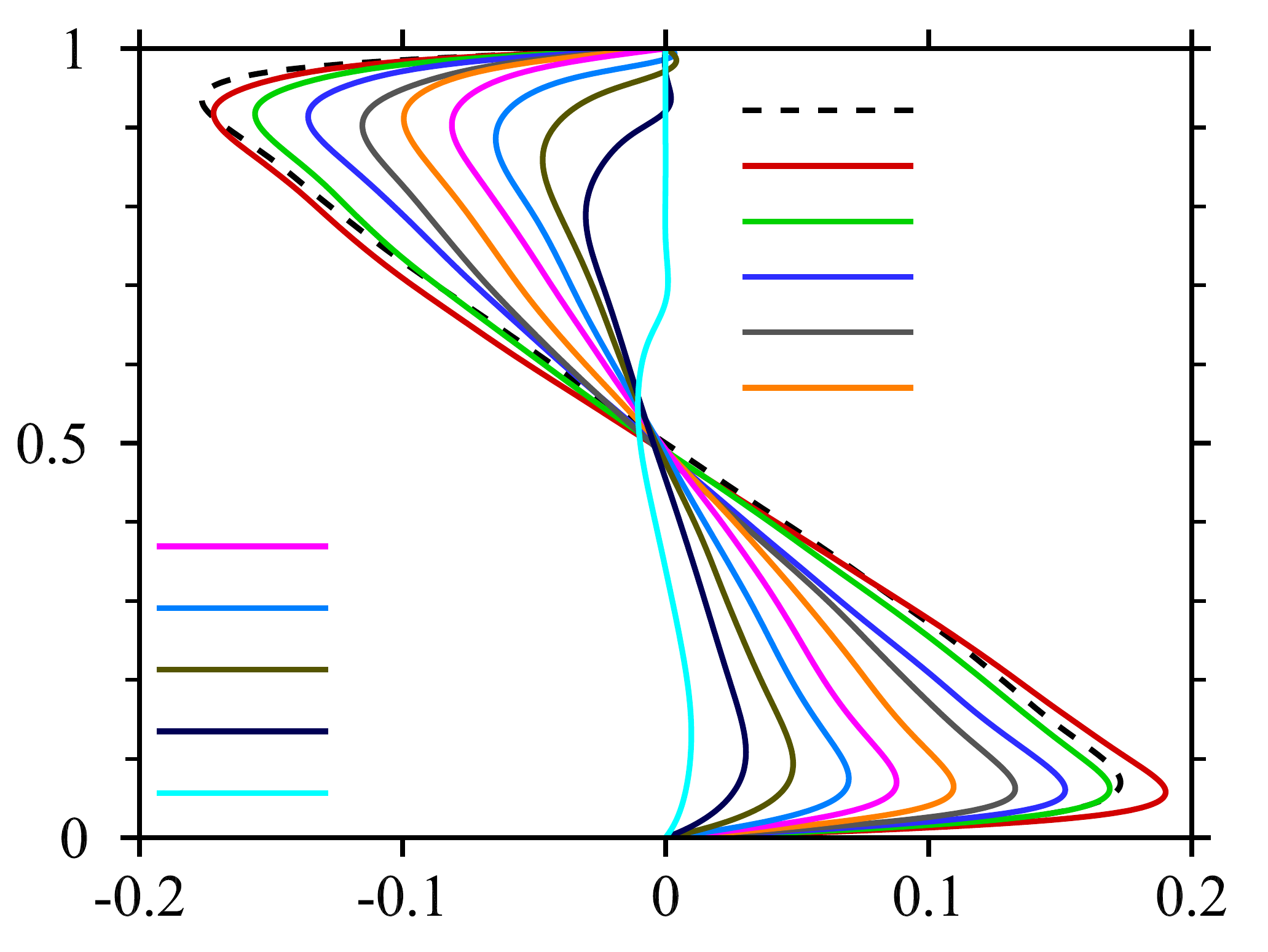}
    \put( -2,71){($a$)}
    \put( -3,40){$z$}
    \put( 52,-1){$v$}
    \put(73,65){\scriptsize{OB}}
       \put(73,61){\scriptsize{$\theta_m=0$}}
       \put(73,56.5){\scriptsize{$\theta_m=0.1$}}
       \put(73,52){\scriptsize{$\theta_m=0.2$}}
       \put(73,48){\scriptsize{$\theta_m=0.3$}}
       \put(73,43.5){\scriptsize{$\theta_m=0.4$}}
        \put(27,31){\scriptsize{$\theta_m=0.5$}}
        \put(27,26){\scriptsize{$\theta_m=0.6$}}
         \put(27,21){\scriptsize{$\theta_m=0.7$}}
          \put(27,16){\scriptsize{$\theta_m=0.8$}}
           \put(27,11.5){\scriptsize{$\theta_m=0.9$}}        
  \end{overpic}
  \begin{overpic}[width=0.45\textwidth]{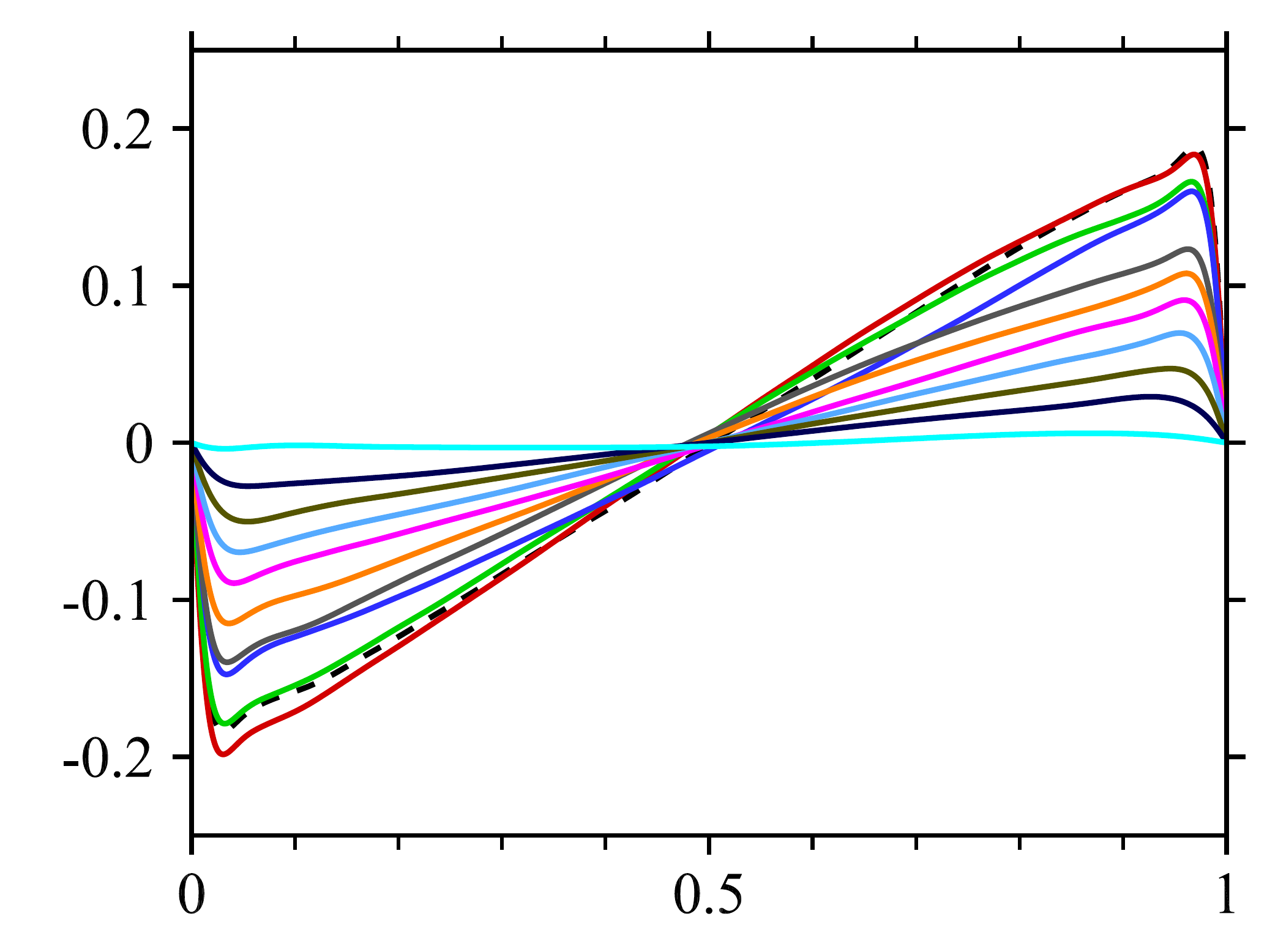}
    \put( 0,71){($b$)}
     \put( -3,40){$w$}
    \put( 54,-1){$y$}
  \end{overpic}
 \caption{Time-averaged velocity profiles for $Ra=10^9$ for 2D cases with $\Gamma=1$ (\textit{a}) $v$ profile at $y=0.5$ for different $\theta_m$. (\textit{b}) $w$ profile at $z=0.5$ for different $\theta_m$.
  }\label{pro}
\end{figure}

We now quantitatively investigate velocity profiles. Figure \ref{pro}($a$) shows time-averaged horizontal velocity $v$ profile at the line $y=0.5$ for 2D cases with $Ra=10^9,\Gamma=1$. The profile is symmetrical for the OB case. For $\theta_m=0$, the profile is very close to the OB case, however, we can see that the peak horizontal velocity near the bottom plate is slightly increased compared to the OB case, while it decreases a little near the top plate. As $\theta_m$ increases, the fluid motion is obviously weakened, and the horizontal velocity peak decreases monotonically with increasing $\theta_m$, the velocity boundary layers that form near the top plate can still be clearly identified for not too large $\theta_m$. As $\theta_m$ increases further to $0.7, 0.8$, there is a region where $v$ is around 0 near the top plate. This corresponds to the stably-stratified flow structure for relative large $\theta_m$.  For $\theta_m=0.9$, there exist flow reversals of the LSC \citep{sugiyama2010flow}, thus we only average counterclockwise circulation for this case. It is seen that convection only happens at the lower layer with $z$ roughly smaller than 0.7. Figure \ref{pro}($b$) shows time-averaged vertical velocity $w$ profile at the line $z=0.5$. For $\theta_m=0$, the profile is also asymmetrical, and the velocity peak is larger than the OB case near the left plate. The vertical velocity peak is also found to decrease with increasing $\theta_m$.

\begin{figure}
  \centering
  \begin{overpic}[width=0.45\textwidth]{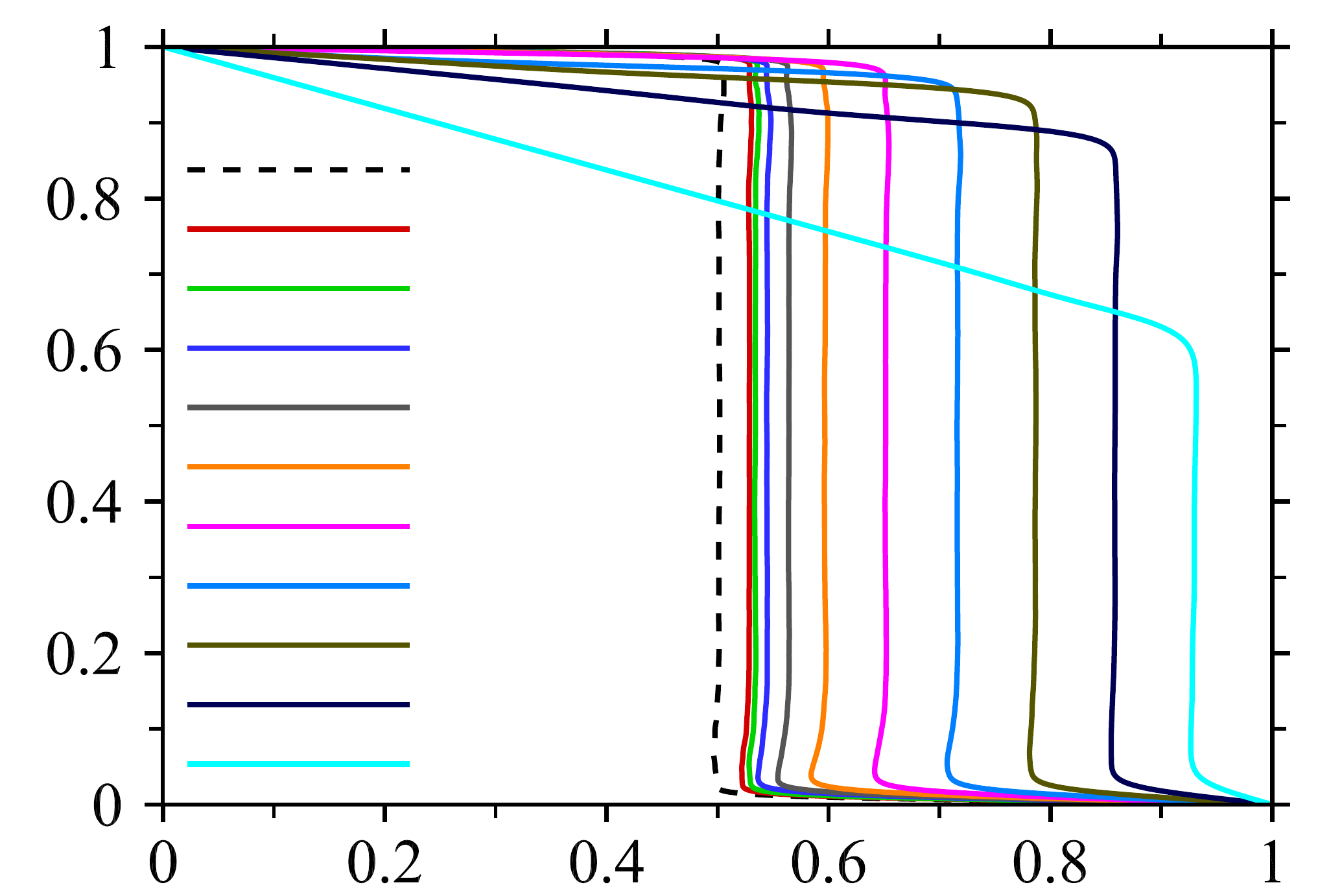}
    \put( -2,63){($a$)}
    \put( -3,34){\large$z$}
      \put(32,53){\scriptsize{OB}}
       \put(32,49){\scriptsize{$\theta_m=0$}}
       \put(32,44){\scriptsize{$\theta_m=0.1$}}
       \put(32,39){\scriptsize{$\theta_m=0.2$}}
       \put(32,35){\scriptsize{$\theta_m=0.3$}}
       \put(32,31){\scriptsize{$\theta_m=0.4$}}
        \put(32,26){\scriptsize{$\theta_m=0.5$}}
        \put(32,22){\scriptsize{$\theta_m=0.6$}}
         \put(32,17.5){\scriptsize{$\theta_m=0.7$}}
          \put(32,13){\scriptsize{$\theta_m=0.8$}}
           \put(32,9){\scriptsize{$\theta_m=0.9$}}           
    \put(52,-4){$\theta$}
  \end{overpic}
  \begin{overpic}[width=0.45\textwidth]{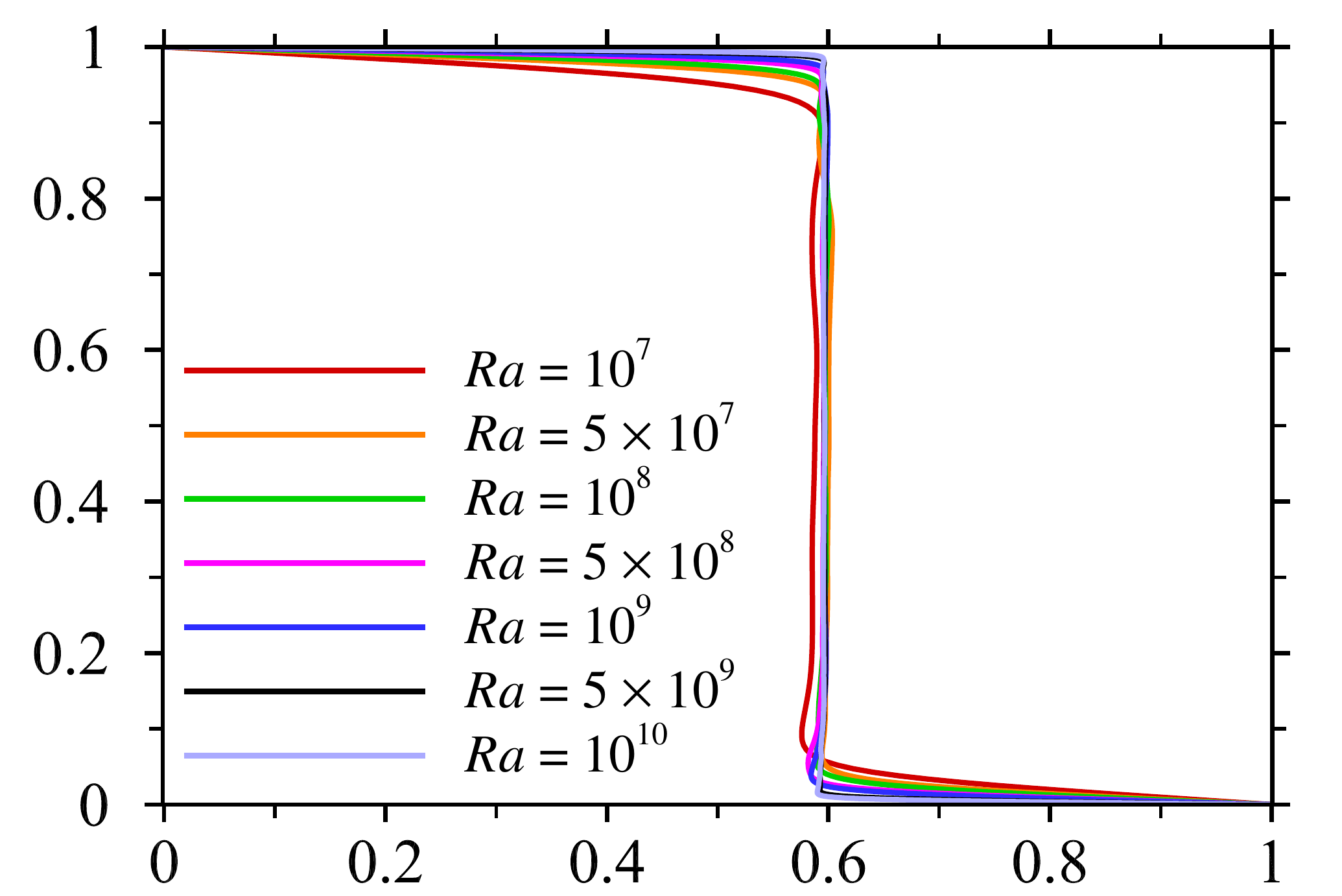}
    \put( -2,63){($b$)}
    \put(52,-4){$\theta$}
  \end{overpic}
  \vskip 2mm
 \caption{Time-averaged temperature profiles at $y=0.5$ for 2D cases with $\Gamma=1$ for  (\textit{$a$})  $Ra=10^9$ with different $\theta_m$ and (\textit{$b$}) $\theta_m$=0.4 with different $Ra$.
  }\label{temp}
\end{figure}
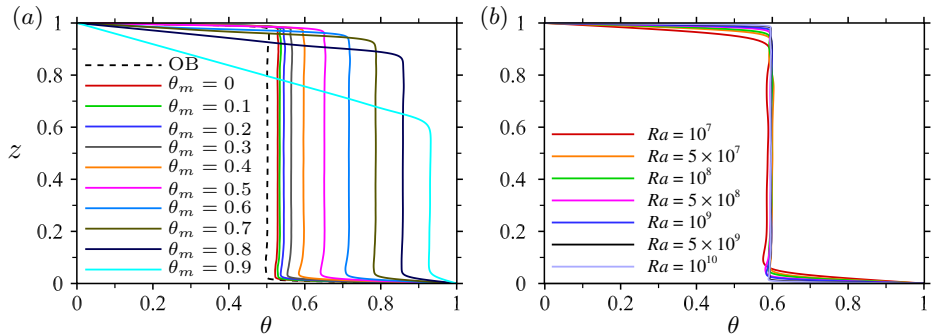

We now investigate thermal boundary layers and centre temperature. In Figure \ref{temp}($a$) we show time-averaged temperature profiles at the location of $y=0.5$ for the 2D case with $Ra=10^9, \Gamma=1$. For the OB case, the temperature at the centre $\theta_c$ is the arithmetic mean of the top- and bottom-plate temperatures. For $\theta_m$=0, it is clearly seen that $\theta_c$ is enhanced compared to that of the OB case.
With increasing $\theta_m$, the thermal BLs become thicker and $\theta_c$ becomes larger. It is also evident that the top BL is thicker than that of the bottom BL. Figure \ref{temp}($b$) shows time-averaged temperature profiles at the location of $y=0.5$ for the 2D cases with $\theta_m=0.4, \Gamma=1$. An interesting feature is that although the thermal BLs become thinner with increasing $Ra$, the bulk temperature still keeps nearly unchanged for all $Ra$.

\begin{figure}
  \centering
  \begin{overpic}[width=0.45\textwidth]{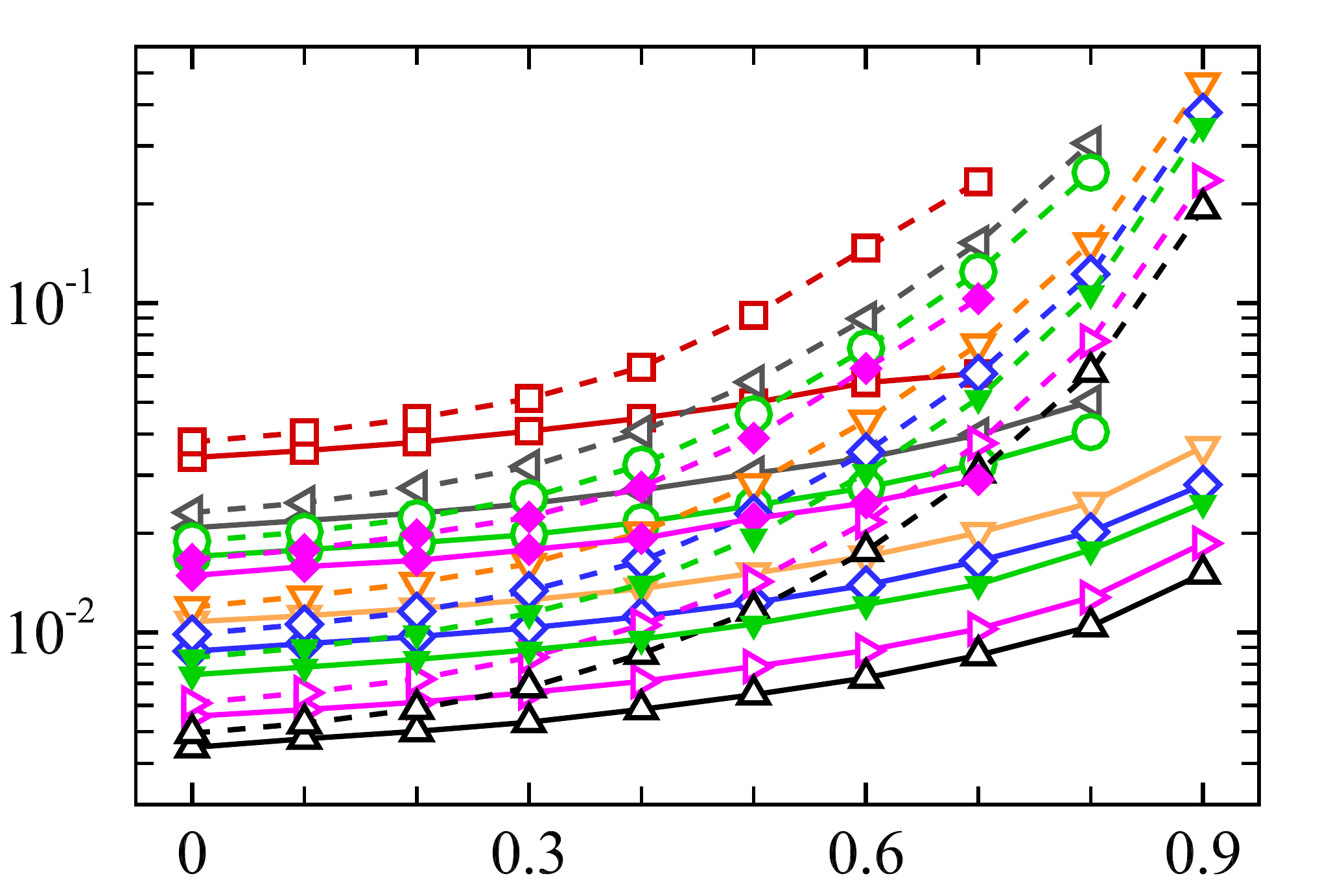}
     \put( -4,61){($a$)}
    \put(52,-3){$\theta_m$}
    \put(-6,28){$\rotatebox{90}{{$\lambda_t^\theta,\lambda_b^\theta$}}$}
  \end{overpic}
\begin{overpic}[width=0.45\textwidth]{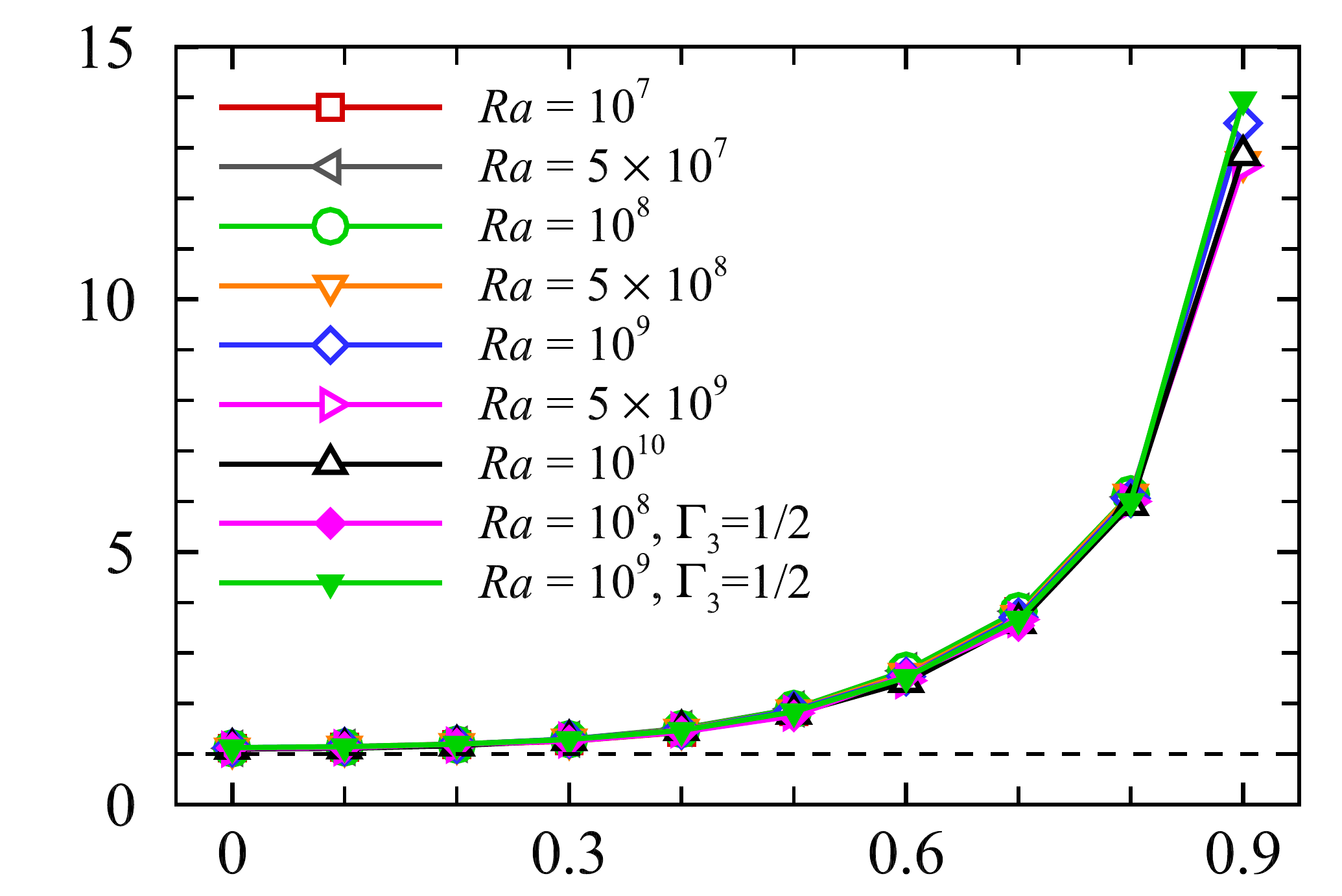}
    \put( -2,61){($b$)}
    \put(-3,35){${F_\lambda}$}
      \put(52,-3){$\theta_m$}
      \end{overpic}
  \vskip 2mm
 \caption{(\textit{a}) Bottom thermal boundary layer thickness $\lambda_b^\theta$ (solid lines) and top  thermal boundary layer thickness $\lambda_t^\theta$ (dashed lines)  as a function of $\theta_m$ for 2D cases with $\Gamma=1$ and 3D cases with $\Gamma_3=1/2$. (\textit{b}) The ratio of the top and bottom thermal BL thicknesses $F_\lambda=\lambda_t^\theta/\lambda_b^\theta$ as a function of $\theta_m$ for  2D cases with $\Gamma=1$ and 3D cases with $\Gamma_3=1/2$.}\label{bl}
\end{figure}

We then quantify the asymmetrical feature of the two thermal BLs.  Figure \ref{bl}($a$) illustrates top and bottom thermal BL thicknesses versus $\theta_m$ for both 2D cases with $\Gamma=1$ and 3D cases with $\Gamma_3=1/2$. The thickness of top thermal BL
$\lambda_t^\theta$ and bottom thermal BL $\lambda_b^\theta$ are defined as \citep{horn2013non}
\begin{gather}
\lambda_t^\theta=\frac{\theta_t-\theta_c}{\partial_z\left<\theta\right>_{A,t}|_{z=1}},~~\lambda_b^\theta=\frac{\theta_c-\theta_b}{\partial_z\left<\theta\right>_{A,t}|_{z=0}}
 \label{eq07}
\end{gather}
 \noindent where $\theta_t=0/\theta_b=1$ are the temperatures at the top/bottom plates. $\left<\right>_{A,t}$ denotes average over any horizontal plane (3D)/line (2D) and in time. One can clearly see that both the two quantities increase with increasing $\theta_m$, and $\lambda_t^\theta$ increases more rapidly. Figure \ref{bl}($b$) shows the ratio of the top and bottom thermal BL thicknesses $F_\lambda=\lambda_t^\theta/\lambda_b^\theta$.  It can be seen  that $F_\lambda$ is an increasing function of $\theta_m$, and it rapidly increases for large $\theta_m$ near 0.9, due to the fact that the stably-stratified flow structures appear for large $\theta_m$, as shown in figures \ref{flow}($c$) and \ref{flow}($f$), leading to a much thicker thermal BL near the top plate as seen in Figure \ref{bl}($a$). One noteworthy feature is that all data sets collapse well on top of each other. It is surprising that $F_\lambda$ has a universal relationship with $\theta_m$ which seems to be independent of $Ra$, and the universality also holds for the 3D case. Note that in the present study we only consider the chaotic cases. For relatively small $Ra$ and large $\theta_m$, chaotic motions disappear, the flow is periodical/quasi-periodical or steady.  A pure conduction state without BLs is even obtained for $Ra=10^7$ and $5\times10^7$ with $\theta_m=0.9$. Previous studies on NOB convection caused by large temperature differences also found that $F_\lambda$  is practically independent of $Ra$, but increases with increasing temperature differences \citep{horn2013non}. Our results demonstrate that this finding also holds for the penetrative RBC, suggesting some similarity between penetrative convection and NOB convection with large temperature differences.

\begin{figure}
  \centering
\begin{overpic}[width=0.7\textwidth]{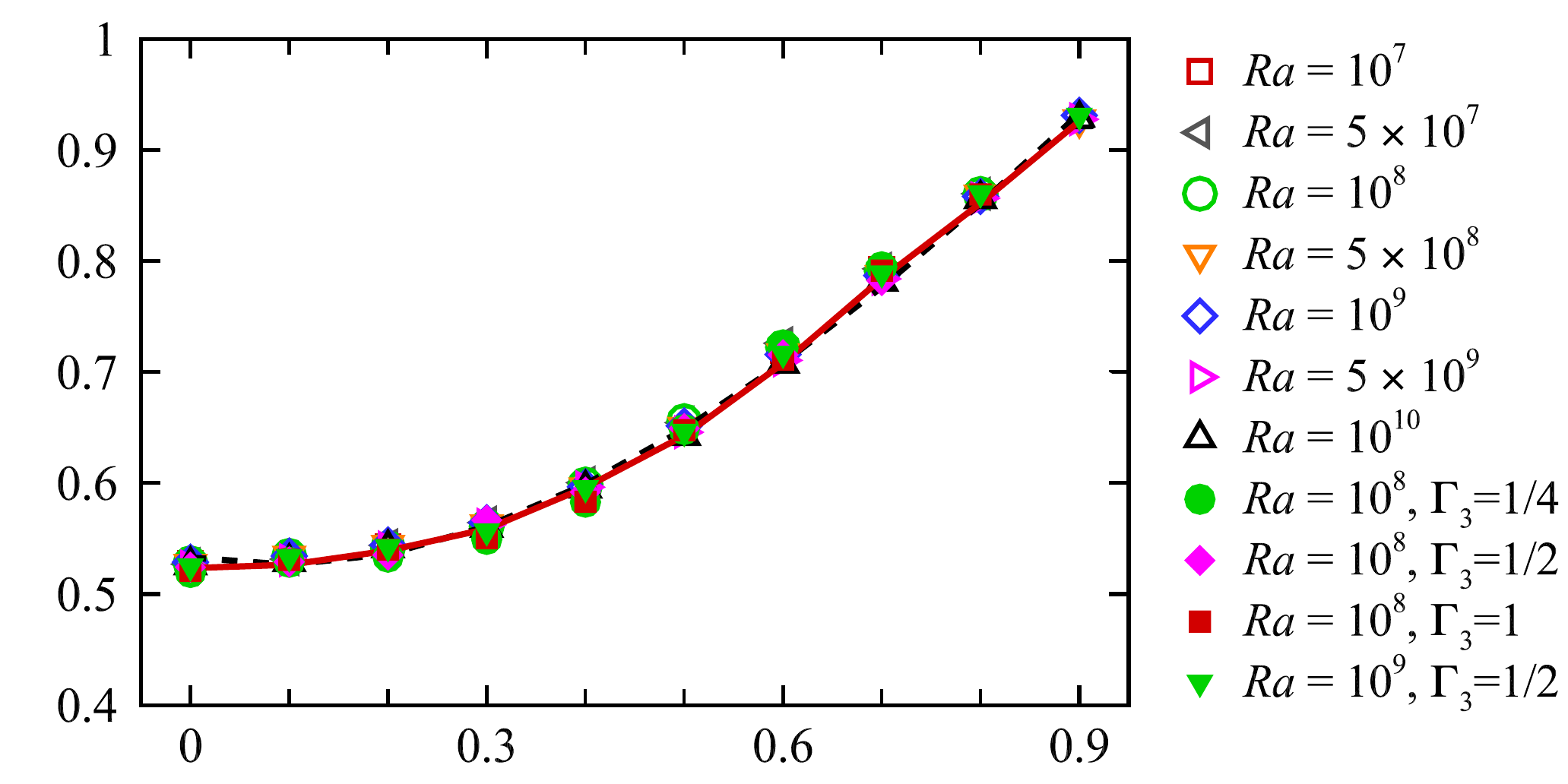}
    \put(-2,26){$\theta_c$}
    \put(40,-2){$\theta_m$}
  \end{overpic}
  \vskip 2mm
 \caption{Centre temperature $\theta_c$ as a function of $\theta_m$ for different $Ra$. Hollow symbols for 2D simulations  with $\Gamma=1$, filled symbols for 3D simulations. The solid line is line segment of ${\lambda_t^\theta}/({\lambda_t^\theta+\lambda_b^\theta})$ calculated at different $\theta_m$ for $Ra=10^{10}$ . The dashed lines is a polynomial fitting which is expressed as $\theta_c=0.530-0.133\theta_m+0.858\theta_m^2-0.237\theta_m^3$.
  }\label{tc}
\end{figure}

The asymmetrical feature of the thermal BLs will lead to a shift of $\theta_c$ compared to the OB cases. In previous NOB studies, it was found that $\theta_c$ increases for water \citep{ahlers2006non,sugiyama2009flow} but decreases for gaseous ethane \citep{ahlers2007non} compared to that of the OB cases. Figure \ref{tc} shows $\theta_c$ as a function of  $\theta_m$ for different $Ra$ for both 2D cases and 3D cases.
An interesting finding is that all the data points  can collapse well onto a single curve. As the thermal conductivity is assumed to be constant, the temperature gradients at the bottom and top walls should be equal, and the $Nu$ can be simply expressed as $Nu=-\partial_z\left<\theta\right>_{A,t}|_{z=1}=-\partial_z\left<\theta\right>_{A,t}|_{z=0}$. Based on equation (\ref{eq07}), one can obtain

\begin{gather}
\theta_c=\frac{\lambda_t^\theta}{\lambda_t^\theta+\lambda_b^\theta}
 \label{tceq}
\end{gather}

\noindent Thus $1/\theta_c=1/F_\lambda+1$. The solid red line in figure \ref{tc} shows line segment of  $\lambda_t^\theta/(\lambda_t^\theta+\lambda_b^\theta)$ as a function of $\theta_m$ calculated at different $\theta_m$  for $Ra=10^{10}$, the data sets for different $Ra$ can well collapse onto this curve. The dashed black curve is a polynomial fitting which is expressed as $\theta_c=0.530-0.133\theta_m+0.858\theta_m^2-0.237\theta_m^3$, the data sets for different $Ra$ can well fall onto this curve.

\subsection{Nusselt number and Reynolds number }

\begin{figure}
  \centering
  \begin{overpic}[width=0.45\textwidth]{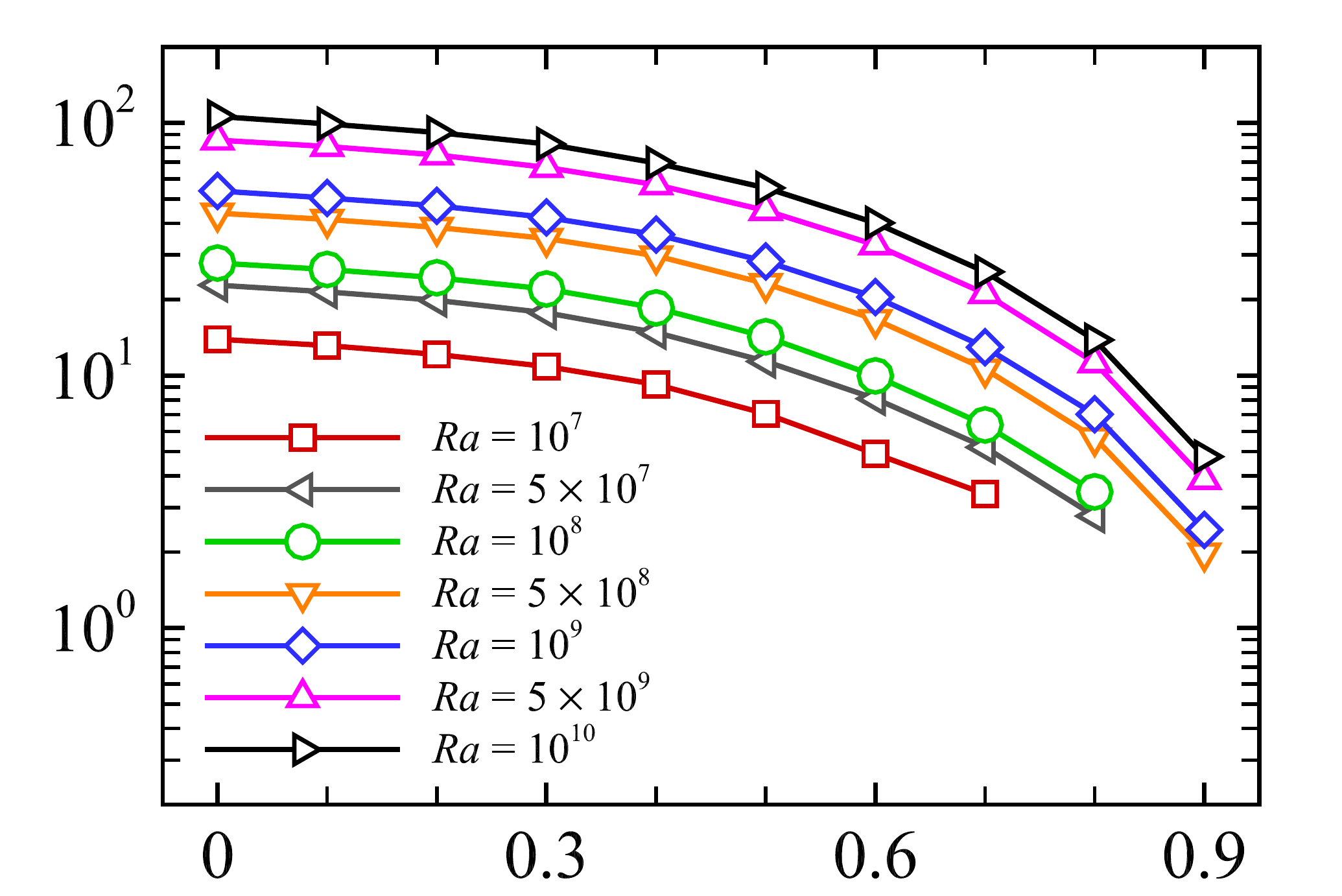}
    \put( -4,62){($a$)}
    \put( -6,35){$Nu$}
  \end{overpic}
  \begin{overpic}[width=0.45\textwidth]{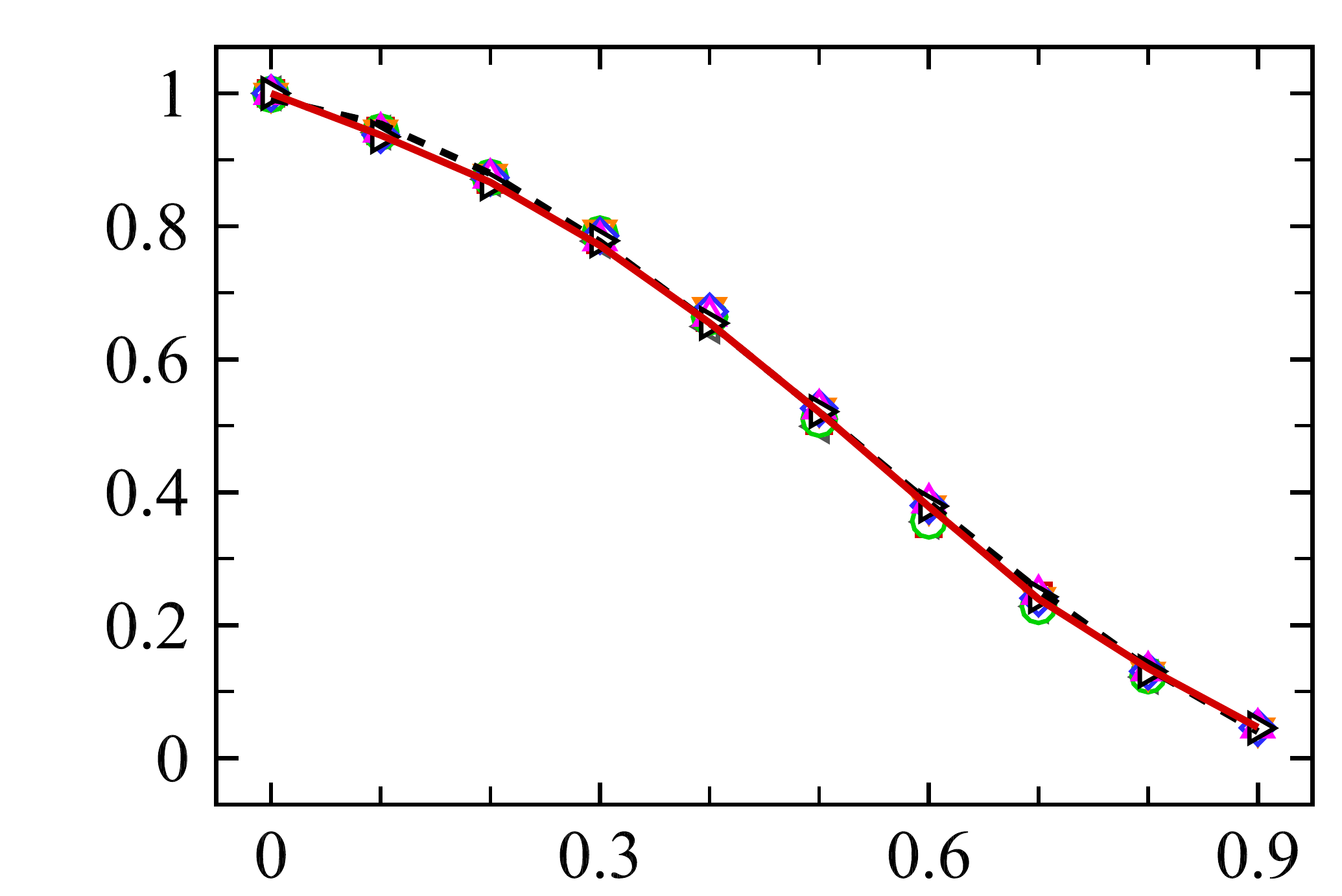}
    \put( -2,62){($b$)}
     \put(-1,20){\rotatebox{90}{$Nu(\theta_m)/Nu(0)$}}
     \end{overpic}
   \begin{overpic}[width=0.45\textwidth]{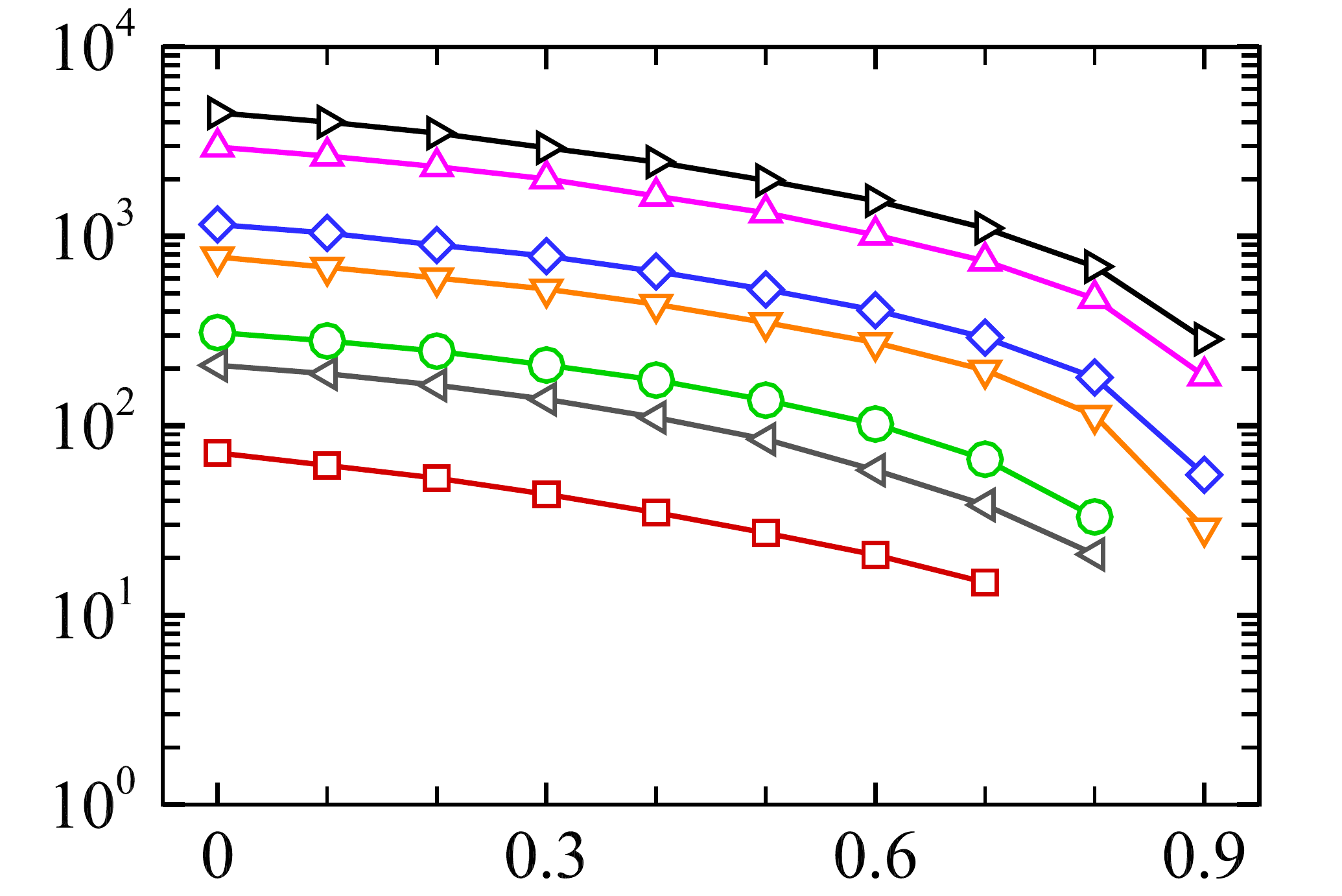}
    \put( -4,62){($c$)}
     \put(-6,35){$Re$}
     \put( 52,-3){$\theta_m$}
  \end{overpic}
  \begin{overpic}[width=0.45\textwidth]{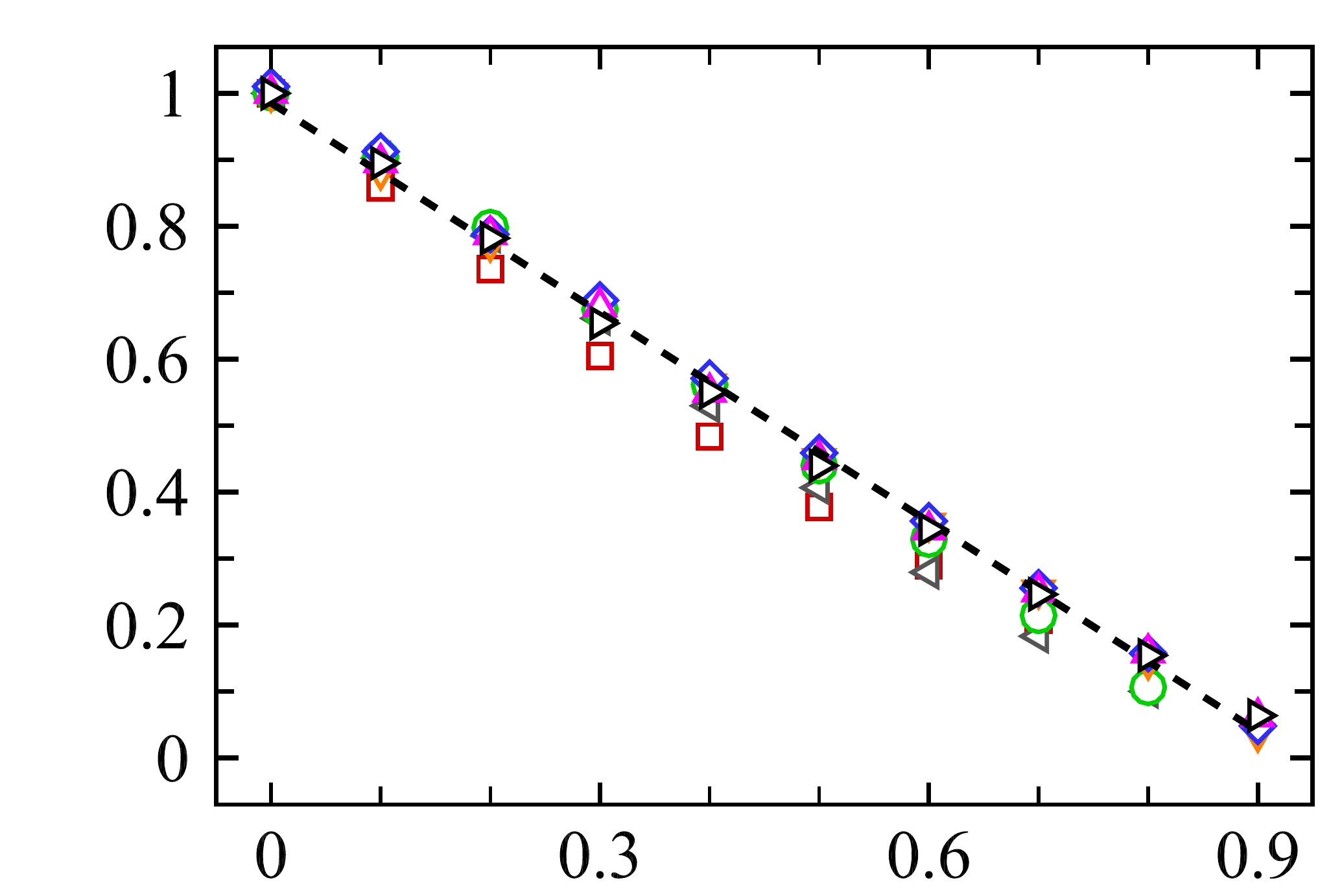}
    \put( -2,62){($d$)}
    \put(-1,20){\rotatebox{90}{$Re(\theta_m)/Re(0)$}}
     \put(52,-3){$\theta_m$}
  \end{overpic}
  \vskip 3mm
 \caption{Absolute and normalized Nusselt numbers and Reynolds numbers as a function of $\theta_m$ for 2D cases with different $Ra$ and fixed $\Gamma=1$ (\textit{a}) Absolute Nusselt numbers $Nu$ (\textit{b}) Normalized Nusselt numbers $Nu(\theta_m)/Nu(0)$, the red sold line denotes line segment of $(\lambda_t^\theta(0)+\lambda_b^\theta(0))/(\lambda_t^\theta(\theta_m)+\lambda_b^\theta(\theta_m))$ calculated for $Ra=10^{10}$ at different $\theta_m$, the black dashed line denotes a polynomial fitting of $Nu(\theta_m)/Nu(0)=0.992-0.170\theta_m-2.244\theta_m^2+1.397\theta_m^3$  (\textit{c}) Absolute Reynolds numbers $Re$ (\textit{d}) Normalized Reynolds numbers $Re(theta_m)/Re(0)$, the black dashed line denotes a linear fitting of $Re(\theta_m)/Re(0)=0.98-1.046\theta_m$.
  }\label{nure}
\end{figure}

We now turn to the heat-transfer properties measured by the Nusselt number $Nu$, and the flow intensity which can be expressed in terms of the Reynolds number $Re$. The $Nu$ and $Re$ are expressed as:

\begin{equation}
Nu=\left<\sqrt{RaPr}u_z\theta-\partial_z\theta\right>_{A,t}, Re=\sqrt{Ra/Pr}U
\end{equation}
where $\left<\right>_{A,t}$ means average over any horizontal plane(3D)/line(2D) and time, $U=\sqrt{(\bold{u}\cdot\bold{u})_{y,z,t}}$. The $Nu$ studied here is averaged over the bottom plate.
In figure \ref{nure}, we show the absolute and normalised $Nu$ and $Re$ as a function of $\theta_m$ for different $Ra$ for 2D cases with $\Gamma=1$. It is seen from figure \ref{nure}($a$) that $Nu$ drops monotonically with increasing $\theta_m$. This is because that with increasing $\theta_m$, the stably-stratified flow regimes gradually appear and grow near the top plate, where the flow is mainly dominated by the thermal conduction. For the extreme limit $\theta_m=1$, the flow should be totally within the conduction regime with $Nu=1$. Figure \ref{nure}($b$) shows the normalized Nusselt number $Nu(\theta_m)/Nu(0)$ as a function of $\theta_m$ for various $Ra$. We can see that all data sets collapse well onto a single curve. From equation (\ref{eq07}), one can obtain
$Nu(\lambda_t^\theta+\lambda_b^\theta)=1$, and thus

\begin{equation}
\frac{Nu(\theta_m)}{Nu(0)}=\frac{\lambda_t^\theta(0)+\lambda_b^\theta(0)}{\lambda_t^\theta(\theta_m)+\lambda_b^\theta(\theta_m)}
\end{equation}

\noindent  We note that this relationship is analytically exact (see also \cite{ahlers2006non,horn2013non}).The red solid line in figure \ref{nure}($b$) shows line segment of $(\lambda_t^\theta(0)+\lambda_b^\theta(0))/(\lambda_t^\theta(\theta_m)+\lambda_b^\theta(\theta_m)$ as a function of $\theta_m$ calculated at different $\theta_m$ for $Ra=10^{10}$. One can see that the data points for other $Ra$ all nicely fall onto this curve. The black dashed curve is a polynomial fitting denoted by $Nu(\theta_m)/Nu(0)=0.992-0.170\theta_m-2.244\theta_m^2+1.397\theta_m^3$, the data sets for different $Ra$ can well collapse on top of this fitted curve. Figure \ref{nure}($c$) shows the absolute $Re$ as a function of $\theta_m$ for different $Ra$. It is seen that $Re$ also decreases monotonically with increasing $\theta_m$. The reduction of $Re$ can be also qualitatively understood, in view that the appearance of stably-stratified flow structures weakens the fluid motion, and in turn drops considerably the magnitude of $Re$. Figure \ref{nure}($d$) plots the normalised Reynolds number $Re(\theta_m)/Re(0)$ as a function of $\theta_m$. One can see that $Re(\theta_m)/Re(0)$ roughly drops linearly with increasing $\theta_m$, and the data sets can be well described by a linear fitting of $Re(\theta_m)/Re(0)=0.980-1.046\theta_m$. The curve is fitted for the $Ra=10^{10}$ data, and only the data sets for relatively low $Ra$ (i.e. $Ra=10^7$ and $5\times10^7$) show some deviations from this fitting.

\begin{figure}
  \centering
  \begin{overpic}[width=0.45\textwidth]{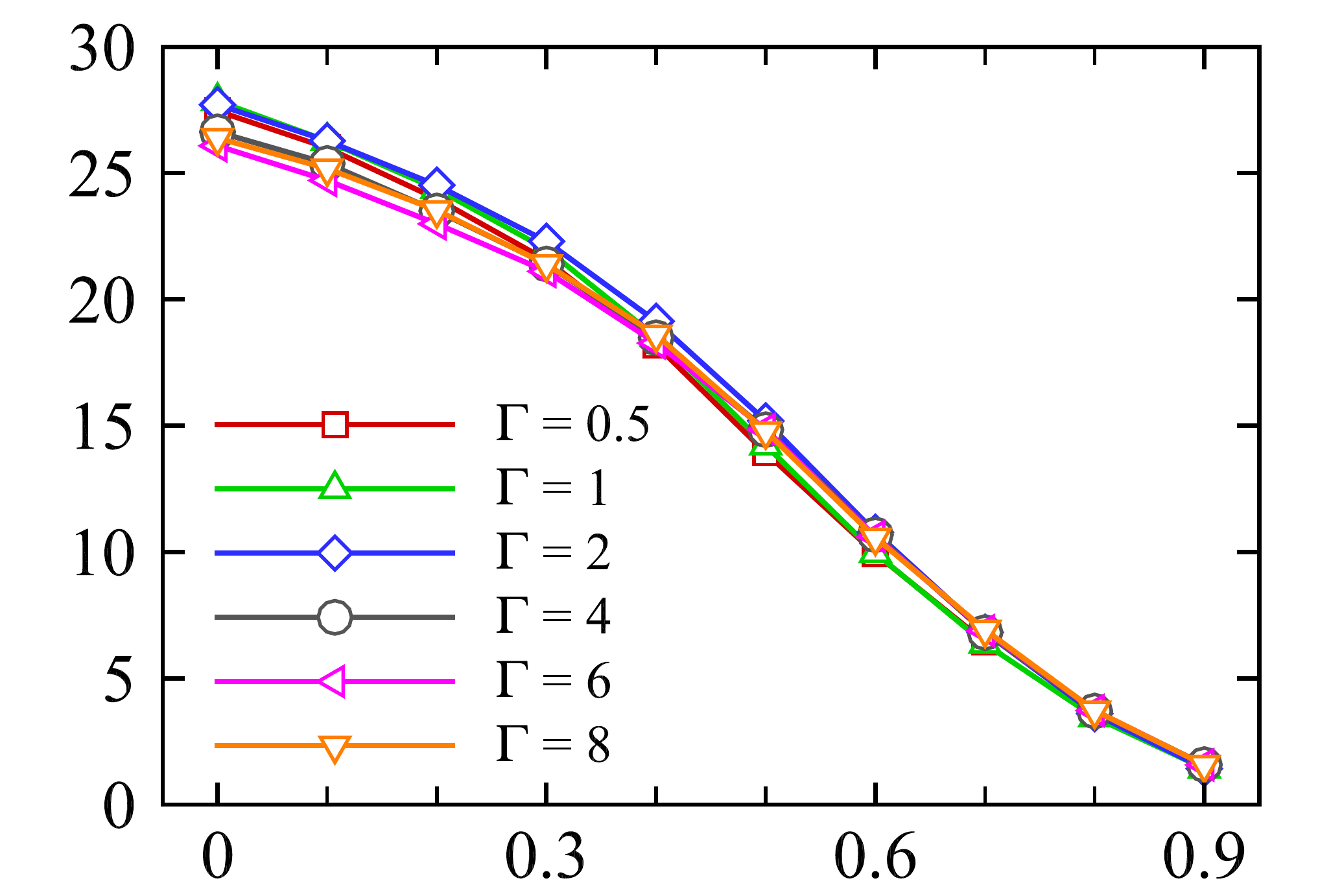}
    \put( -5,62){($a$)}
    \put( -6,35){$Nu$}
  \end{overpic}
  \begin{overpic}[width=0.45\textwidth]{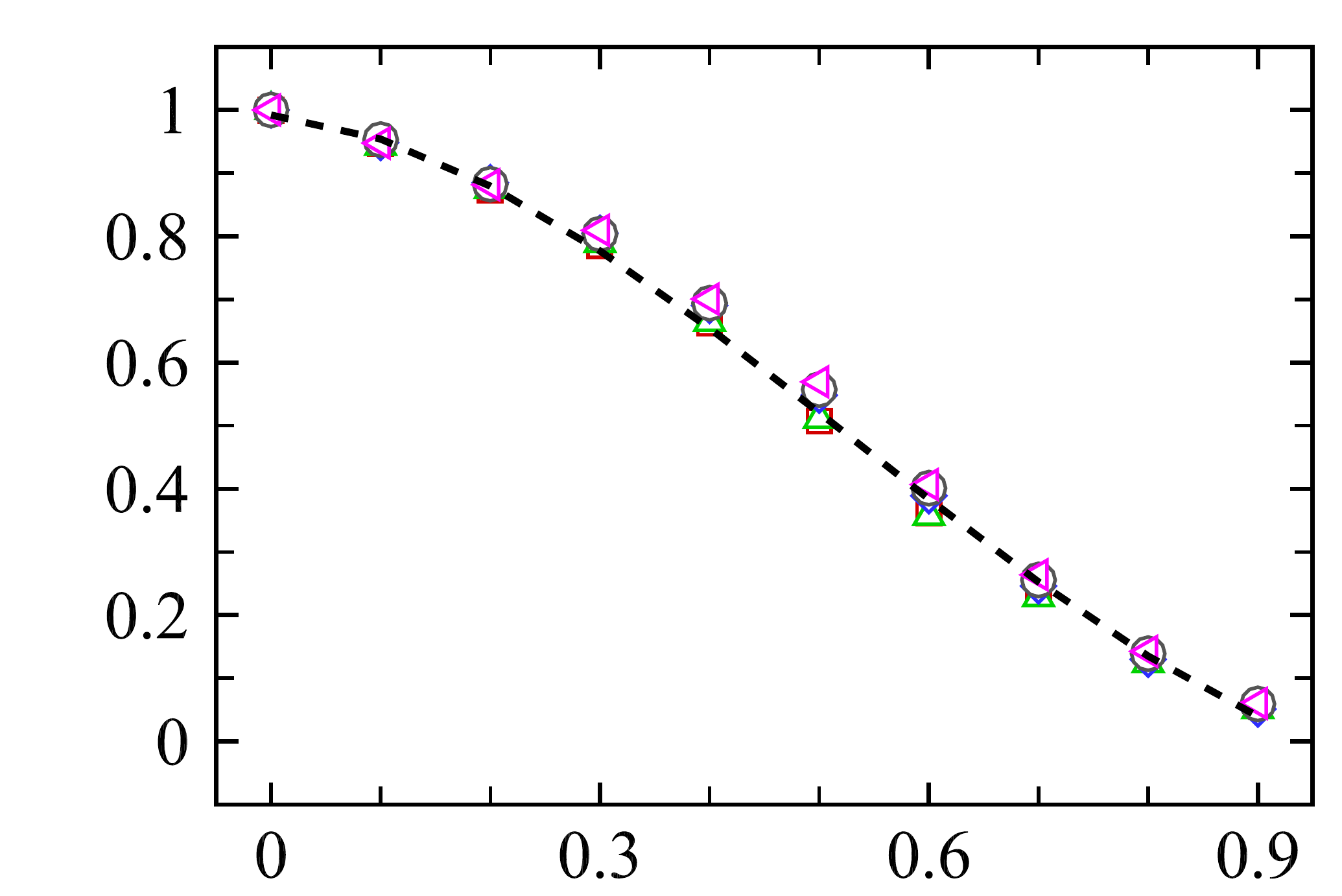}
    \put( 0,62){($b$)}
     \put(-2,20){\rotatebox{90}{$Nu(\textcolor{red}{\theta_m})/Nu(0)$}}
  \end{overpic}
   \begin{overpic}[width=0.45\textwidth]{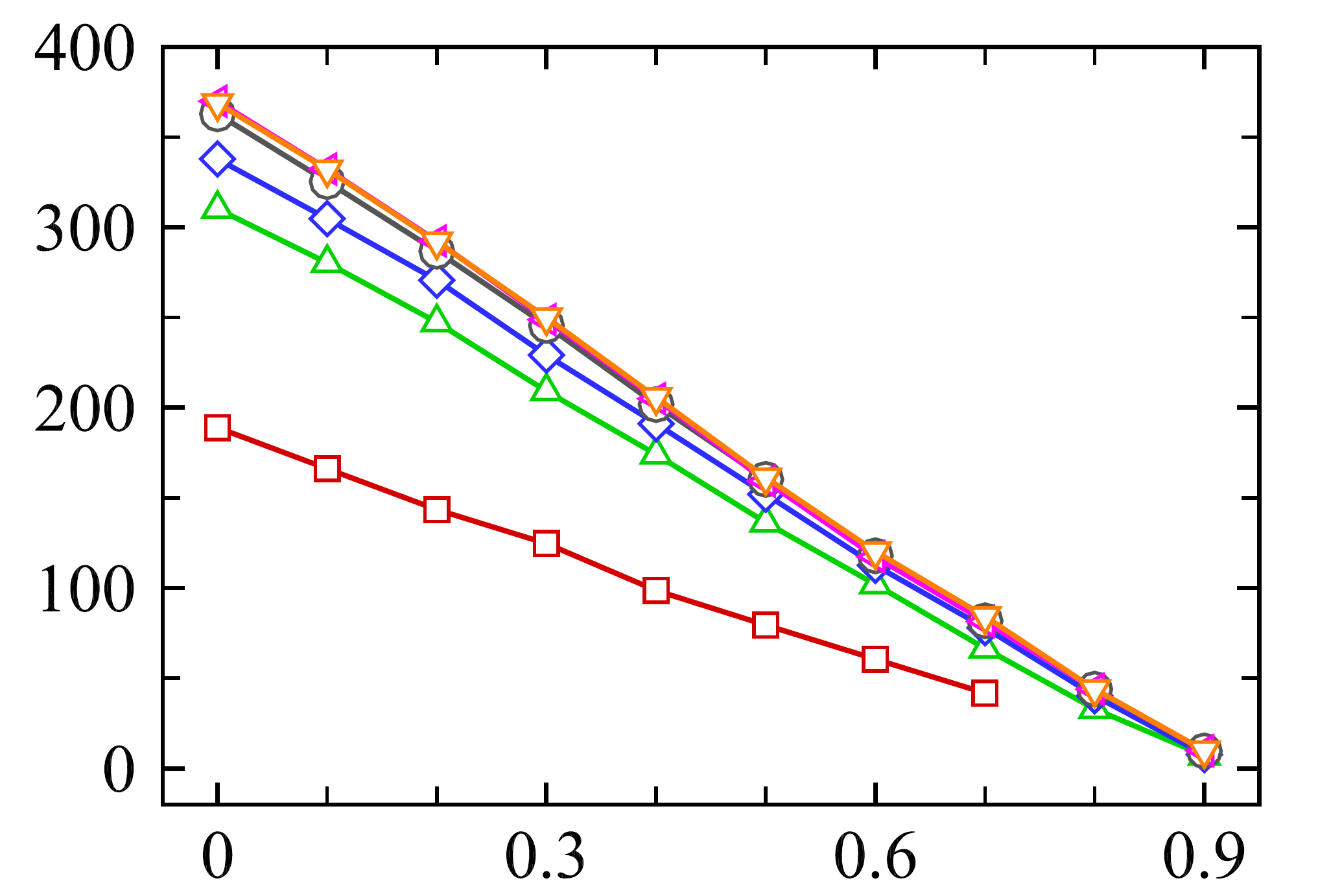}
    \put( -5,62){($c$)}
     \put(-6,35){$Re$}
     \put( 52,-3){$\theta_m$}
  \end{overpic}
  \begin{overpic}[width=0.45\textwidth]{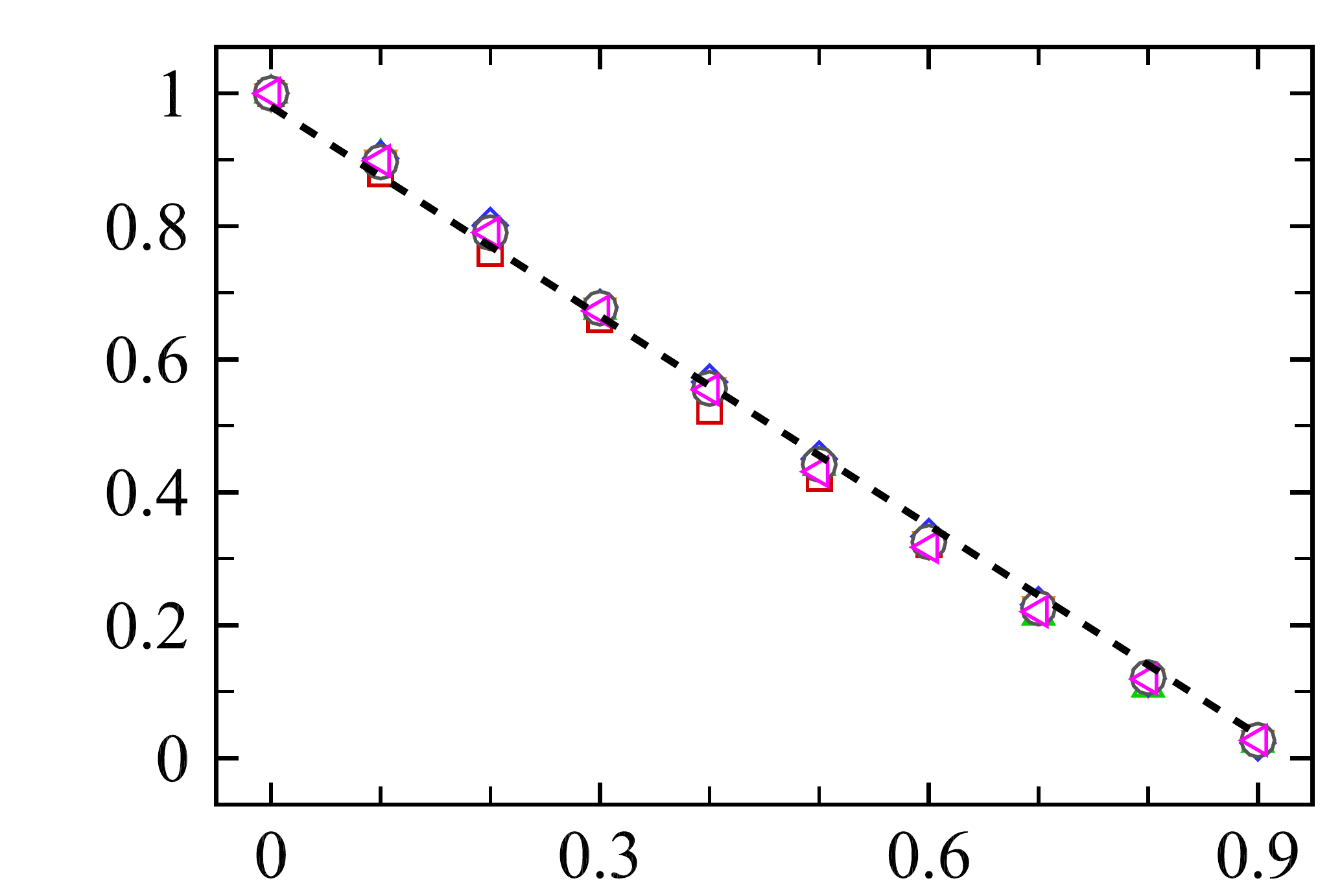}
    \put(0,62){($d$)}
    \put(-2,20){\rotatebox{90}{$Re(\textcolor{red}{\theta_m})/Re(0)$}}
     \put(52,-3){$\theta_m$}
  \end{overpic}
  \vskip 3mm
 \caption{Absolute and normalized Nusselt numbers and Reynolds numbers as a function of $\theta_m$ for different $\Gamma$ and fixed $Ra=10^8$ for 2D simulations. (\textit{a}) Absolute Nusselt numbers $Nu$. (\textit{b}) Normalized Nusselt numbers $Nu(\theta_m)/Nu(0)$. (\textit{c}) Absolute Reynolds numbers $Re$. (\textit{d}) Normalized Reynolds numbers $Re(\theta_m)/Re(0)$. The dashed lines are the same as figure \ref{nure}.
  }\label{nurear}
\end{figure}

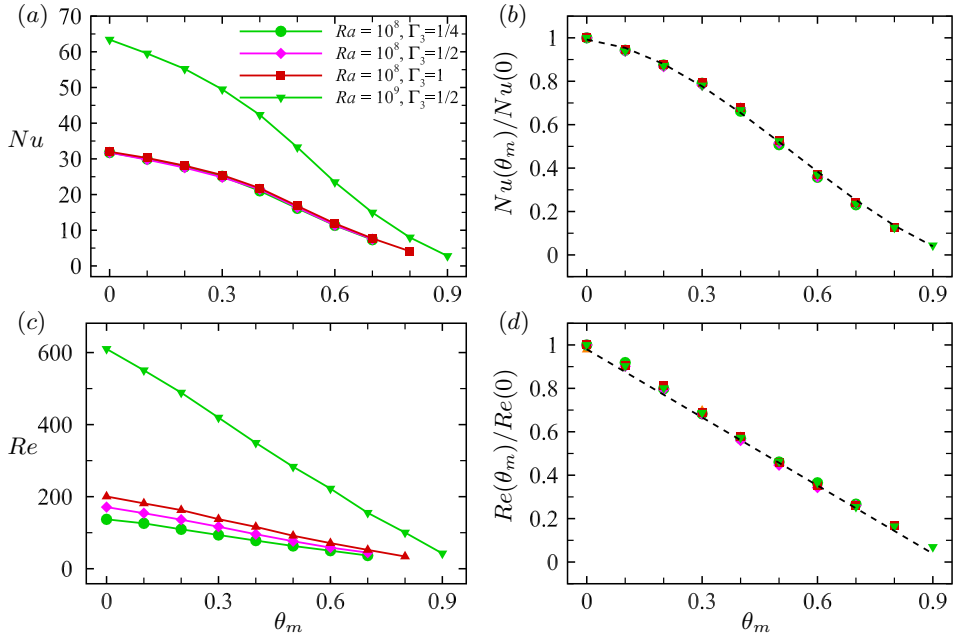
\begin{figure}
  \centering
  \begin{overpic}[width=0.45\textwidth]{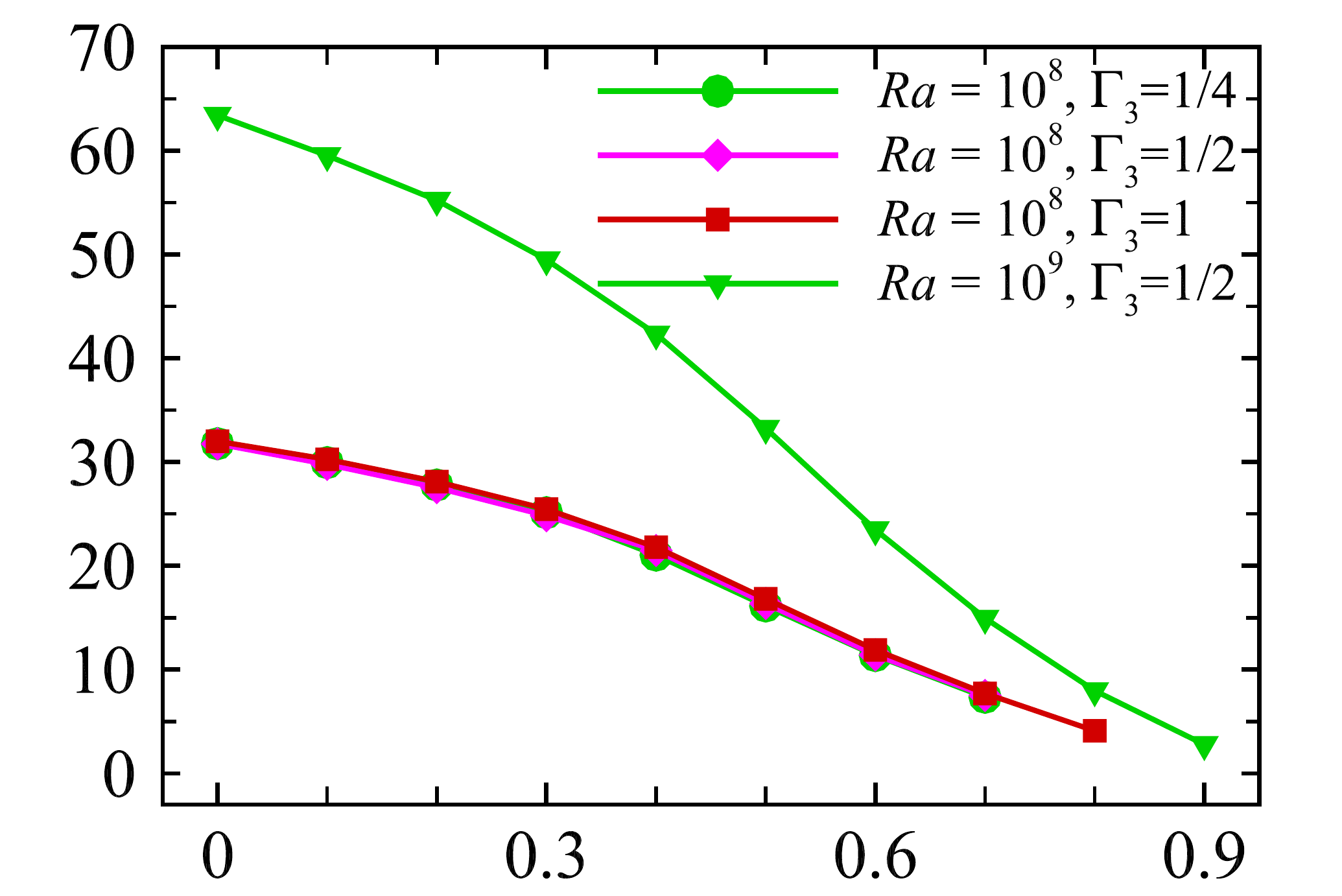}
    \put( -4,62){($a$)}
    \put( -6,35){$Nu$}
  \end{overpic}
  \begin{overpic}[width=0.45\textwidth]{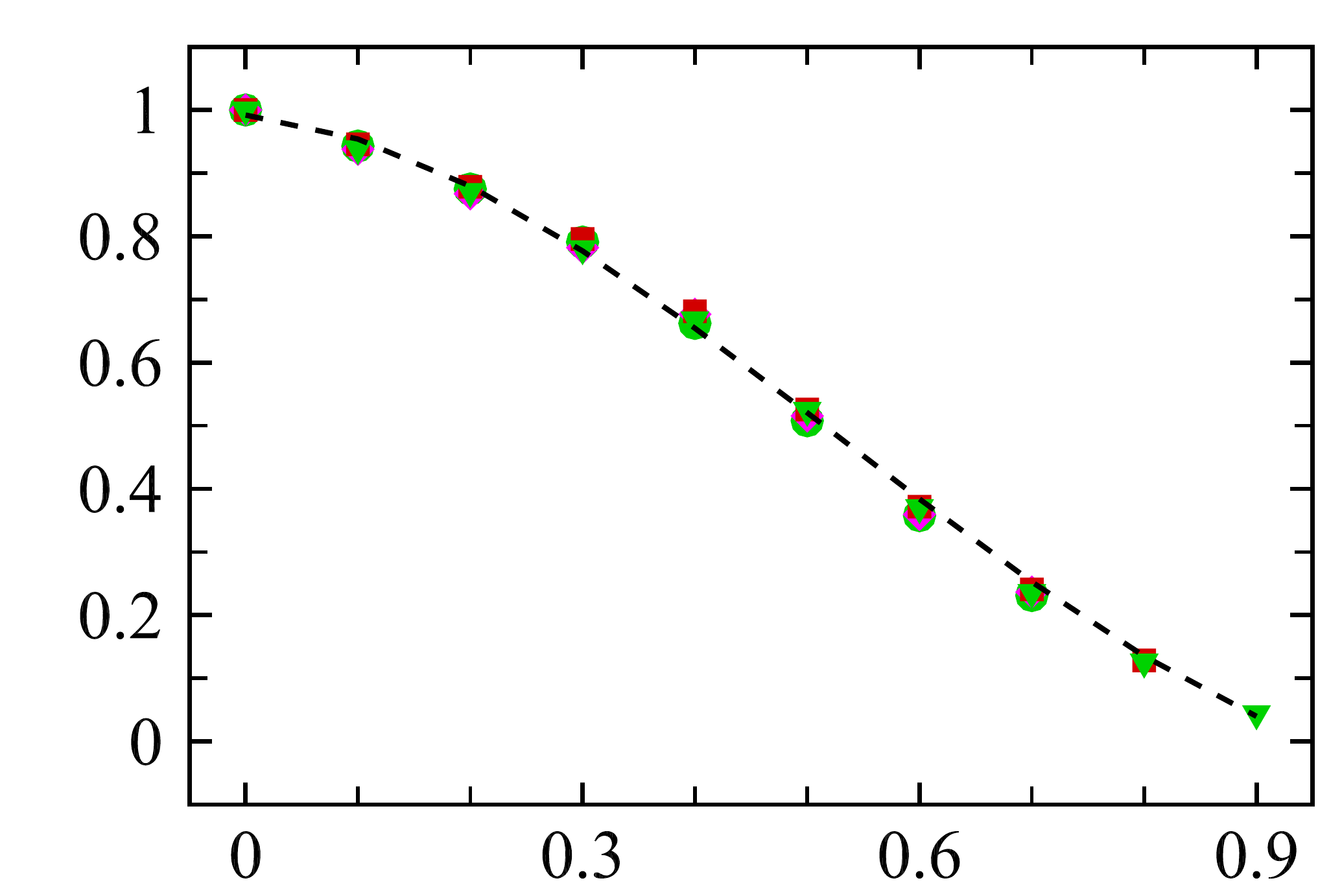}
    \put( -1,62){($b$)}
     \put(-2,20){\rotatebox{90}{$Nu(\theta_m)/Nu(0)$}}
  \end{overpic}
   \begin{overpic}[width=0.45\textwidth]{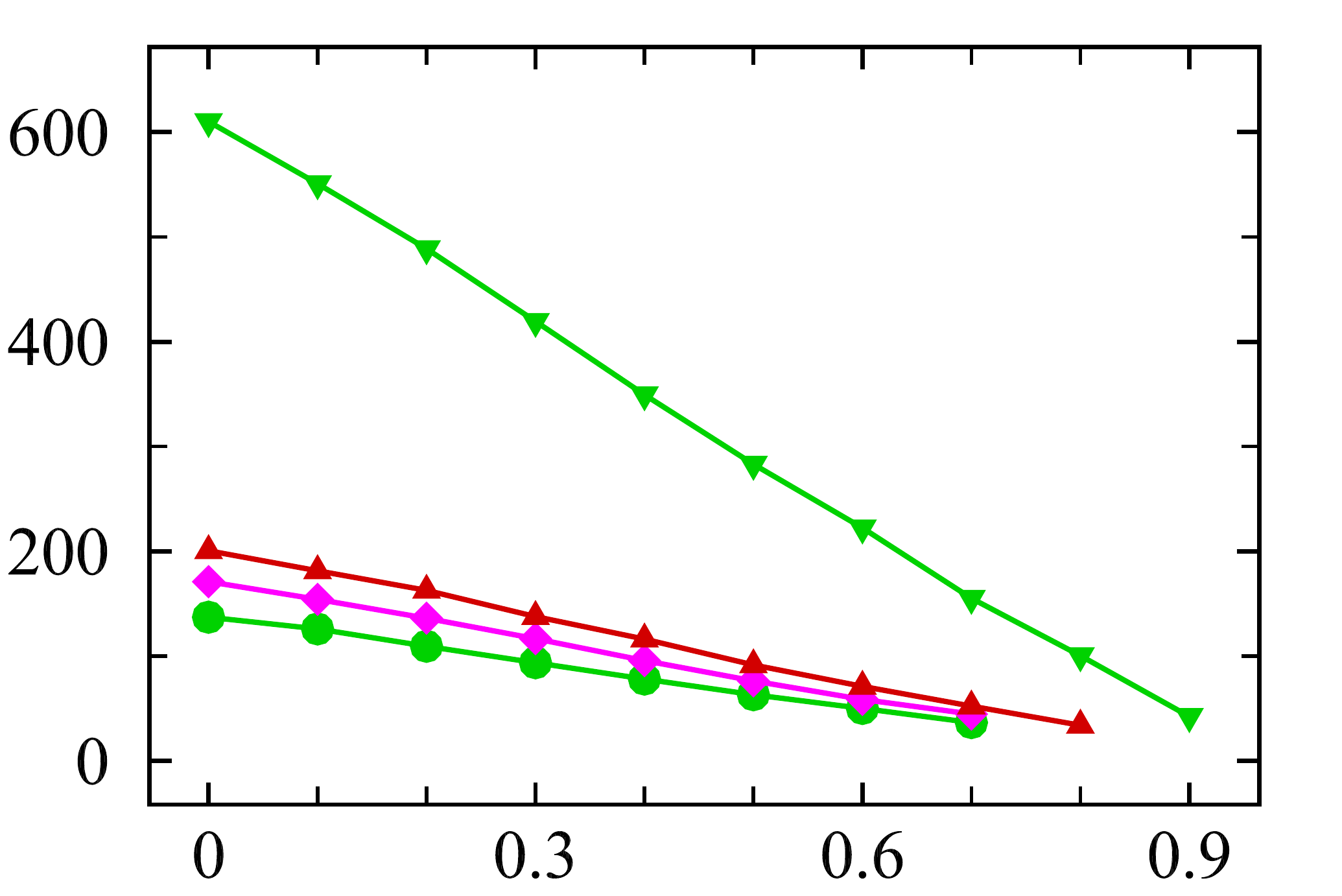}
    \put( -4,62){($c$)}
     \put(-6,35){$Re$}
     \put( 52,-3){$\theta_m$}
  \end{overpic}
  \begin{overpic}[width=0.45\textwidth]{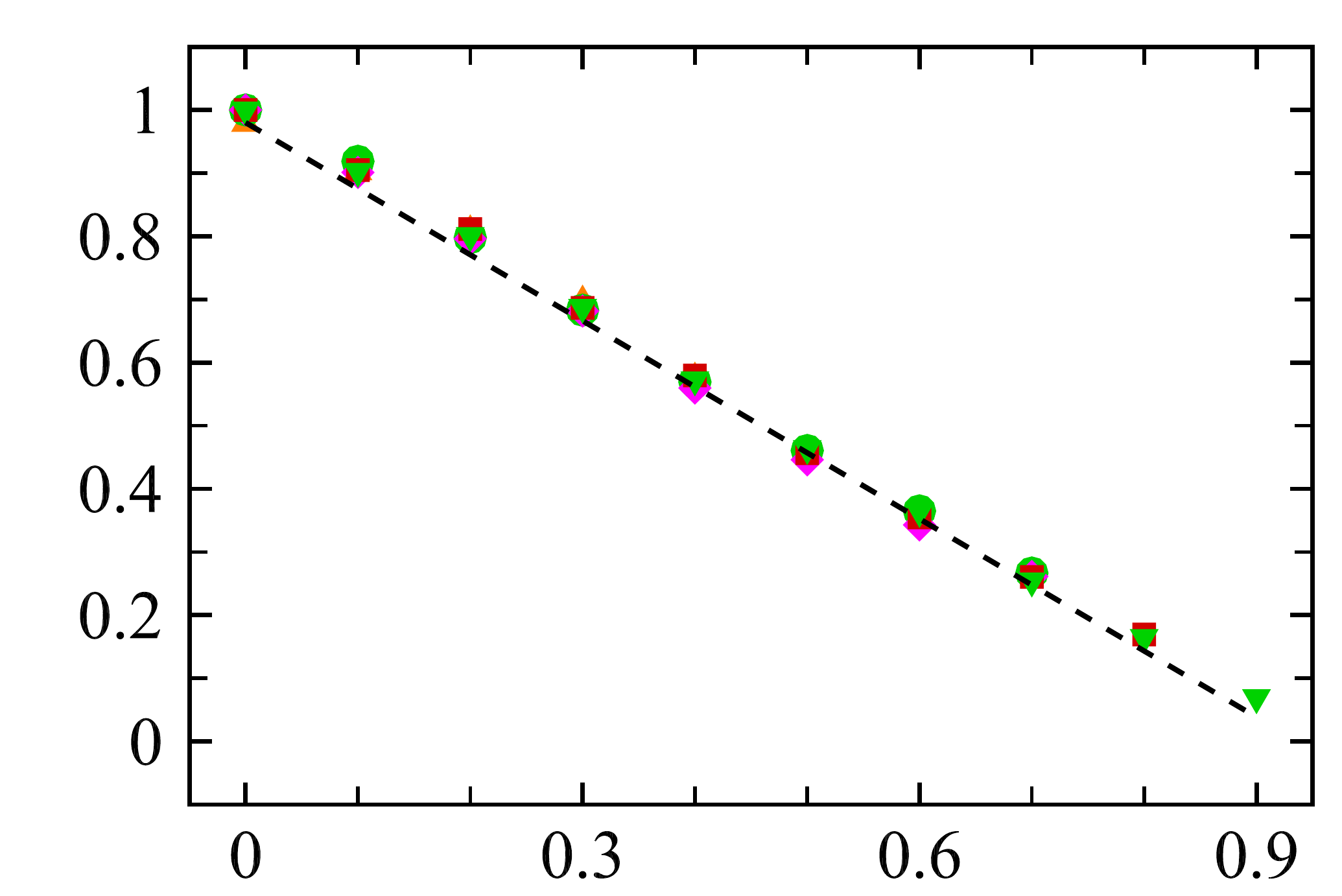}
    \put( -1,62){($d$)}
    \put(-2,20){\rotatebox{90}{$Re(\theta_m)/Re(0)$}}
     \put(52,-3){$\theta_m$}
  \end{overpic}
  \vskip 3mm
 \caption{Absolute and normalized Nusselt numbers and Reynolds numbers as a function of $\theta_m$ for 3D cases with different $\Gamma_3$. (\textit{a}) Absolute Nusselt numbers $Nu$ (\textit{b}) Normalized Nusselt numbers $Nu(\theta_m)/Nu(0)$. (\textit{c}) Absolute Reynolds numbers $Re$ (\textit{d}) Normalized Reynolds numbers $Re(\theta_m)/Re(0)$. The dashed lines are the same as figure \ref{nure}. }\label{nure3d}
\end{figure}

We then investigate the aspect ratio $\Gamma$ dependence of $Nu$ and $Re$. Figure \ref{nurear} shows the absolute and normalised $Nu$ and $Re$ as a function of $\theta_m$ for different $\Gamma$ and fixed $Ra=10^8$ for 2D simulations. For large $\Gamma$, it was already found that multiple states with different numbers of convection rolls may exist \citep{wang2018multiple}, and this will lead to different heat transfer properties. To avoid such possible multiple states from complicating the present problem, we gradually increase $\theta_m$ using the flow field for smaller $\theta_m$ as initial conditions. The $Nu$ still reduces monotonically with increasing $\theta_m$ and again the normalized Nusselt numbers for different $\Gamma$ can collapse well on top of each other, implying the insensitivity of this dependence on $\Gamma$.  From figure \ref{nurear}($c$), one sees that for all considered $\Gamma$, $Re$ decreases linearly with increasing $\theta_m$, and drops with decreasing $\Gamma$ due to the geometric confinement of the sidewalls. The normalized Reynolds number $Re(\theta_m)/Re(0)$ for different $\Gamma$ also collapse well onto the fitted curve $Re(\theta_m)/Re(0)=0.980-1.046\theta_m$ as shown in figure \ref{nurear}($d$), suggesting an aspect ratio insensitive feature.

 We also show the 3D results in figure \ref{nure3d} in order to find whether the universal behaviour for normalised Nusselt number and Reynolds number found in 2D can be extended to 3D cases. For $Ra=10^8$, we only consider the cases with $\Gamma_3\ge1/4$. For smaller $\Gamma_3$ where the flow in the confined \citep{huang2013confinement,chong2015condensation,chong2017confined,chong2018effect} or severely confined regime \citep{chong2016exploring}, the flow is laminar for a wide range of $\theta_m$, and it is not the focus of the present study. We can clearly see in figure \ref{nure3d}($a$) that the absolute $Nu$ is less influenced by $\Gamma_3$ for all considered $\theta_m$, despite the fact that the flow strength is weakened when reducing $\Gamma_3$ as illustrated in figure \ref{nure3d}($c$). After normalisation, the data for both $Nu$ and $Re$ again fall onto a single curve as seen in figures \ref{nure3d}($b$) and  \ref{nure3d}($d$). Thus, we can conclude that the universal behaviour for normalised $Nu$ and $Re$ can also be found in 3D configurations.

\begin{figure}
  \centering
  \begin{overpic}[width=0.45\textwidth]{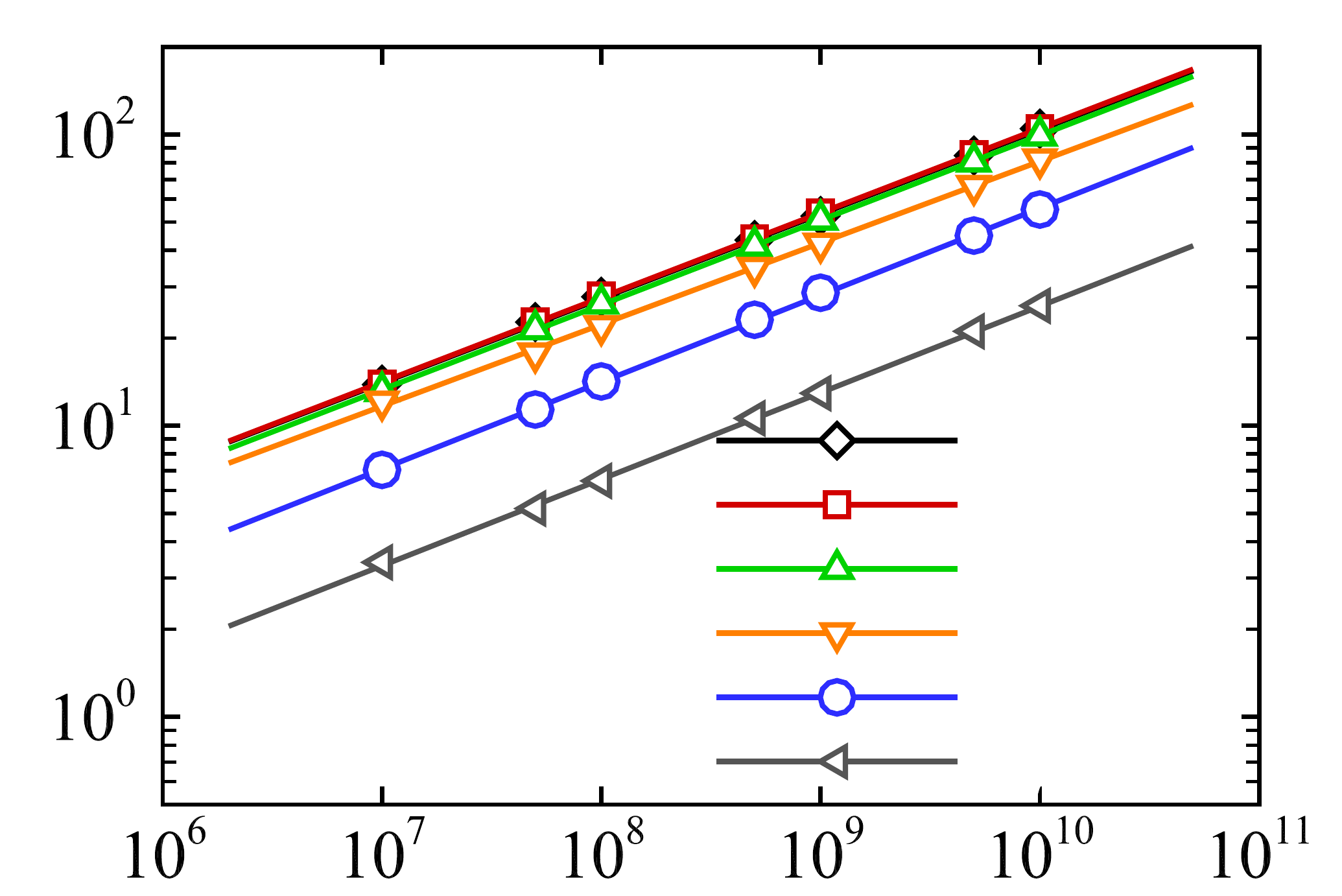}
    \put( -4,62){($a$)}
    \put( -6,35){$Nu$}
    \put(74,33){\scriptsize{OB}}
    \put(74,28){\scriptsize{$\theta_m=0$}}
    \put(74,23){\scriptsize{$\theta_m=0.1$}}
    \put(74,18.5){\scriptsize{$\theta_m=0.3$}}
    \put(74,14){\scriptsize{$\theta_m=0.5$}}
    \put(74,9){\scriptsize{$\theta_m=0.7$}}
     \put(52,-4){$Ra$}
  \end{overpic}
  \begin{overpic}[width=0.45\textwidth]{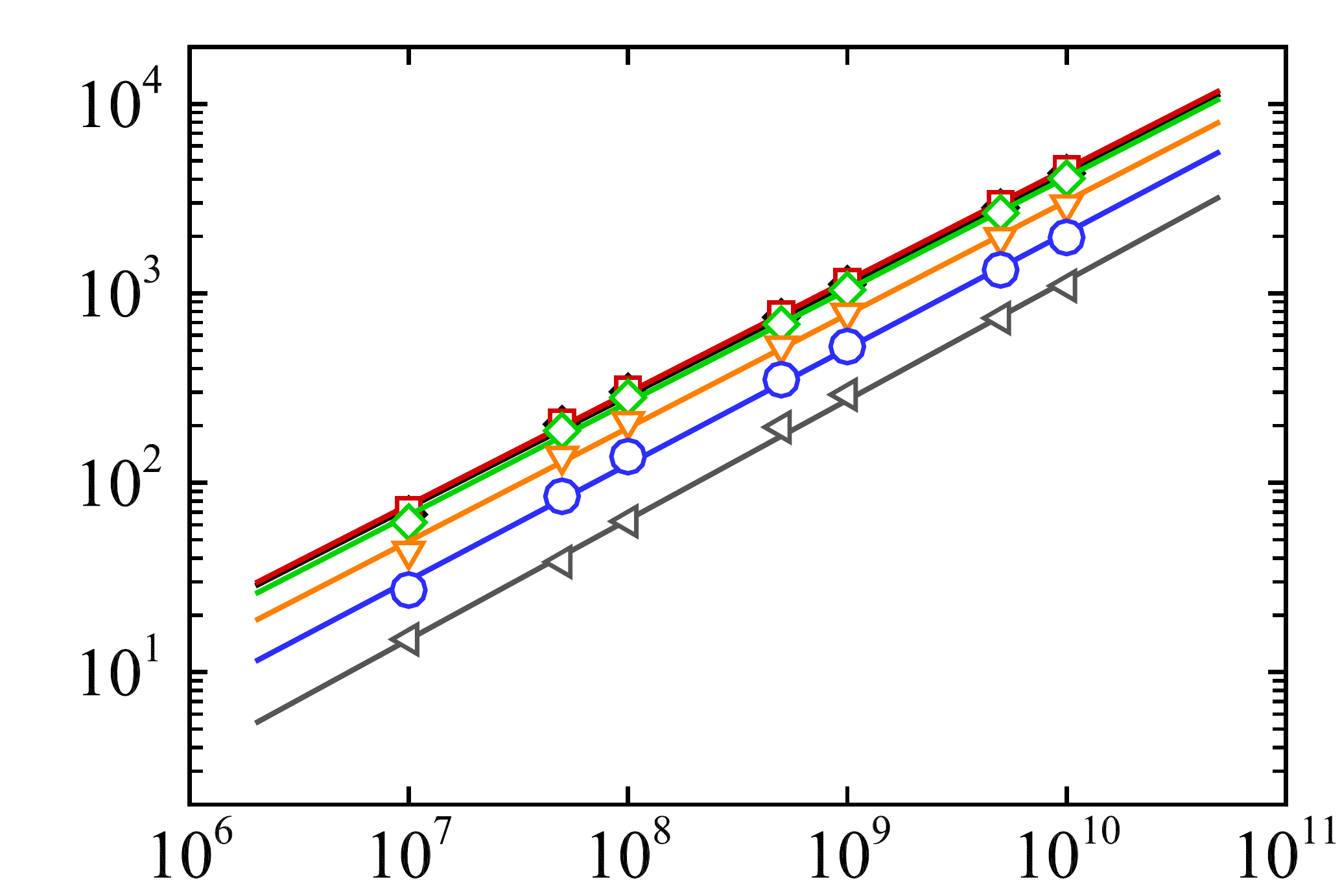}
    \put( -2,62){($b$)}
    \put( -5,35){$Re$}
     \put(52,-4){$Ra$}
  \end{overpic}
\vskip 2mm
 \caption{(\textit{a}) Nusselt numbers $Nu$  and (\textit{b})  Reynolds numbers $Re$ as a function of $Ra$ for different $\theta_m$ for 2D simulations.
  }\label{scaling}
\end{figure}
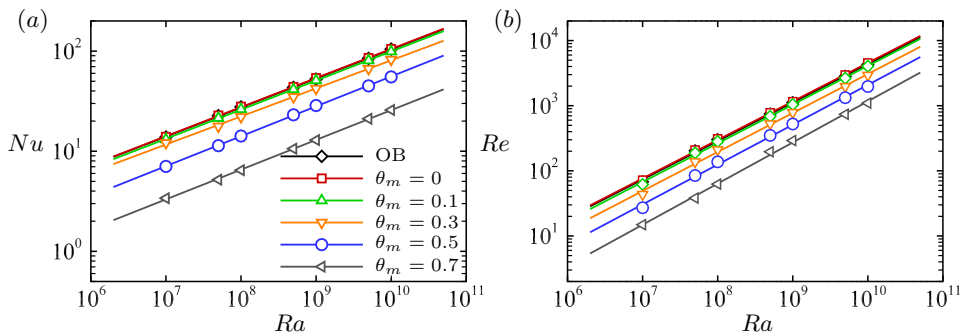

\begin{table}
\begin{center}
\def~{\hphantom{0}}
\begin{tabular}{p{1.5cm}p{1.5cm}p{1.5cm}p{1.5cm}p{1.5cm}}
$\theta_m$ & $a$ & $\alpha$ & $b$ & $\beta$ \\
OB   & 0.131 & 0.290 & 5.43e-3 & 0.590 \\
0   & 0.131 & 0.291 & 5.53e-3 & 0.591 \\
0.1 & 0.123 & 0.291 &4.67e-3 & 0.594 \\
0.2 & 0.113 & 0.291 & 3.94e-3 & 0.596 \\
0.3 & 0.128 & 0.280 & 3.17e-3 & 0.599\\
0.4 & 0.086 & 0.291 & 2.26e-3 & 0.606 \\
0.5 & 0.058 & 0.299 & 1.59e-3 & 0.612 \\
0.6 & 0.036 & 0.305 & 1.01e-3 & 0.621 \\
0.7 & 0.028 & 0.297 & 5.62e-4 & 0.632 \\
0.8 & 0.013 & 0.302 & 1.60e-4 & 0.667\\
0.9 & 6.18e-3 & 0.288 & 1.77e-5& 0.722 \\
\end{tabular}
\caption{The coefficients of the fitted scalings $Nu \sim aRa^{\alpha}$ and $Re \sim bRa^{\beta}$ }\label{tb01}
\end{center}
\end{table}

We finally discuss the scalings of the $Nu$ and $Re$ versus $Ra$ for different $\theta_m$. As mentioned above $Nu(\theta_m)/Nu(0)$ and $Re(\theta_m)/Re(0)$ both have a universal relationship with $\theta_m$ which are independent of $Ra$, and thus the scaling exponents of $Nu\sim a Ra^\alpha$ and $Re \sim b Ra^\beta$ for different $\theta_m$ should also be expected to be independent of $\theta_m$. Figure \ref{scaling}($a$) shows a log-log plot of $Nu$ versus $Ra$ for the OB and penetrative cases with different $\theta_m$ for 2D simulations. Detailed exponents and prefactors for all the $\theta_m$ considered in this work are listed in table \ref{tb01}. For the OB case, the scaling exponent 0.29 is quite close to those obtained in the previous 2D studies. For example, \cite{johnston2009comparison} obtained 0.285 for periodical cells with $Pr=1$, \cite{huang2013counter} obtained 0.29 for $Pr=10$ for no-slip sidewalls similar to the present study. \cite{zhang2017statistics} obtained $0.3\pm0.02$ for $Pr=0.71$ and 5.3 also for no-slip sidewalls. For penetrative cases all the exponents vary between 0.28 and 0.3 for different $\theta_m$, i.e. the scaling exponent does not change much for the penetrative cases. Figure \ref{scaling}($b$) shows a log-log plot of $Re$ vs. $Ra$ for 2D simulations. For the OB case, the scaling exponent 0.59 is also quite close to previous 2D studies (0.62 for $Pr=4.3$ \citep{sugiyama2009flow} and 0.6 for $Pr=0.71$ and 5.3 \citep{zhang2017statistics}.)  For penetrative cases as listed in table \ref{tb01}, one can see that the scaling exponent for $Re$ only deviates from the OB case for large $\theta_m$ close to $0.9$.  According to the data listed in table \ref{tb01}, one may come to the conclusion that the scaling exponents almost keep unchanged for the penetrative cases for not too large $\theta_m$. It is interesting that the scaling exponents are so robust to this density maximum effect despite of the remarkable change of the flow organisations as mentioned before.

\section{Conclusion} \label{sec4}

In summary, we have  studied penetrative turbulent RBC
of water near $4^\circ\rm{C}$ using 2D and 3D DNSs. In the present system, the flow dynamics and heat transport  highly depend on an important parameter, i.e. the density inversion parameter $\theta_m$. Several universal properties are identified and are found to hold over a wide range of parameters for both 2D and 3D. The ratio of the top and bottom thermal BL thickness $F_\lambda$ increases with increasing $\theta_m$, and the relationship is found to be independent of $Ra$.
Similar to the previous studies for NOB convection caused by large temperature differences,
the centre temperature $\theta_c$ is particularly investigated. $\theta_c$ is found to be increased compared to that of the OB cases. As $\theta_c$ is related to $F_\lambda$ with $1/\theta_c=1/F_\lambda+1$, $\theta_c$ also has a universal relationship with $\theta_m$ which is valid for both 2D and 3D cases. With increasing $\theta_m$, both the heat transfer and the flow intensity are substantially weakened.
The normalized Nusselt number $Nu(\theta_m)/Nu(0)$ and Reynolds number $Re(\theta_m)/Re(0)$ also have universal dependence on $\theta_m$ which seems to be independent of both $Ra$ and the aspect ratio $\Gamma$. Finally, the scaling laws of $Nu$ and $Re$ versus $Ra$ are investigated.
It is interesting to find that the scaling exponents of $Nu\sim Ra^\alpha$ and $Re\sim Ra^\beta$, i.e. $\alpha \approx 0.29$ and $\beta \approx 0.59$, are insensitive to $\theta_m$, despite the drastic changes in the flow organizations.

\section*{Appendix. Simulation details} \label{app}

The details of the main 2D simulations are listed in table \ref{tab2d} and the 3D simulations in table \ref{sim3d}. In order not to make the tables too long, we only list some of the main cases for 2D simulations. It can be seen that the differences between $Nu$ at the hot and cold plates are generally within $1\%$. 

\begin{table}
\tabcolsep 6pt
\renewcommand{\arraystretch}{1.}
\label{table1}
\begin{center}
\begin{tabular}{ccccccccccc}
$Ra$      &      $\theta_m$   & $Ny\times Nz $    &  $N_b$   & $N_t$  & $Nu_b$ &  $Re$ &  $\theta_c$  & 100$\Delta Nu$ &$t_{avg}$ \\
$10^7$   &      OB                &    $128\times 128$                   &  10             &    10        &  13.83 & 68.00 &  0.4995  & 0.29   & 6000  \\
$10^7$   &      0                &    $128\times 128$                   &      9         &  10          &  13.92 & 71.86 &  0.5277 &  0.12& 6000   \\
$10^7$   &      0.1                &    $128\times 128$                   &    9           &  11         &  13.13 & 61.69 &  0.5320 &  0.11& 6000   \\
$10^7$   &      0.2                &    $128\times 128$                   &      9         &  11          &  12.11 & 52.86 &  0.5425 &  0.17& 6000   \\
$10^7$   &      0.3                &    $128\times 128$                   &    11           &   13         &  10.85 & 43.47 &  0.5571& 0.17 & 6000  \\
$10^7$   &      0.4                &    $128\times 128$                   &    11          &   15         &  9.21 & 34.77 &  0.5880 &  0.03& 6000   \\
$10^7$   &      0.5                &    $128\times 128$                   &     13          &  20          &  7.05 & 27.13 &  0.6488 &0.17 & 6000  \\
$10^7$   &      0.6                &    $128\times 128$                   &     14          &   29         &  4.89 & 20.78 &  0.7195 &  0.15& 6000   \\
$10^7$   &      0.7                &    $128\times 128$                   &     15          &   39         &  3.39 & 14.89 &  0.7923  &0.14& 5000  \\
$       $    &                         &                                                 &               &            &   &  &   &  \\
$10^8$   &      OB                &    $256\times 256$                   &  10             & 10           &  27.70 & 302.06 &  0.5032 & 0.13 & 3000  \\
$10^8$   &      0                   &    $256\times 256$                   &   10            & 11            &  27.86 & 310.17 &  0.5279 & 0.14 & 3000  \\
$10^8$   &      0.1                   &    $256\times 256$                   &  10             &   11          &  26.27 & 280.26 &  0.5345 & 0.24 & 3000  \\
$10^8$   &      0.2                   &    $256\times 256$                   &    11           &    12        &  24.37 & 247.35 &  0.5415 & 0.14 & 3000  \\
$10^8$   &      0.3                   &    $256\times 256$                   &  11             &   14         & 21.99 & 209.11 &  0.5609&0.26  & 3000  \\
$10^8$   &      0.4                   &    $256\times 256$                   &  12             &   17        & 18.51 & 174.06 &  0.5988 &0.54  & 3000  \\
$10^8$   &      0.5                  &    $256\times 256$                   &     13          &   23         &  14.22 & 136.34 &  0.6559 & 0.14 & 3000  \\
$10^8$   &      0.6                   &    $256\times 256$                   &     15          &    34        & 9.96 & 102.07 &  0.7224&0.04  & 3000  \\
$10^8$   &      0.7                   &    $256\times 256$                   &     17          &    51        &  6.37 & 66.52 &  0.7909 & 0.19& 3000  \\
$10^8$   &      0.8                   &    $256\times 256$                   &     21          &   81         & 3.43 & 33.38 &  0.8606&0.64  & 3000  \\
$       $    &                         &                                                 &               &            &   &  &   &  \\
$10^9$   &      OB                &    $512\times 512$                   &   11            &     11       &  53.31 & 1110.99 &  0.4994&  0.12& 1000  \\
$10^9$   &      0                   &    $512\times 512$                   &   11            &      12      &  53.77 & 1152.52 &  0.5287&0.35 & 1000  \\
$10^9$   &      0.1                   &    $512\times 512$                   &     11          &    12        &  50.41 & 1040.75 &  0.5342&0.15 & 1000  \\
$10^9$   &      0.2                   &    $512\times 512$                   &    11           &     14       &  46.94 & 898.78 &  0.5443&0.14 & 1000  \\
$10^9$   &      0.3                   &    $512\times 512$                   &   12            &   16         & 42.23 & 785.55 &  0.5643& 0.07 & 1600  \\
$10^9$   &      0.4                   &    $512\times 512$                   &     13         &     19       &  36.11 & 651.50 &  0.5967&0.39 & 1600  \\
$10^9$   &      0.5                  &    $512\times 512$                   &     14          &   26         &  28.28 & 523.61 &  0.6516 & 0.47& 1600  \\
$10^9$   &      0.6                   &    $512\times 512$                   &    16          &    37        &  20.42 & 406.66 &  0.7159&0.62 & 1600  \\
$10^9$   &      0.7                   &    $512\times 512$                   &   19            &   59         &  12.93 & 291.71 &  0.7870 &0.31 & 2000  \\
$10^9$   &      0.8                   &    $512\times 512$                   &     23         &   101         &  7.02 & 179.57 &  0.5287&0.05 & 2000  \\
$10^9$   &      0.9                   &    $256\times 256$                   &     15          &  107          &  2.45 & 55.09 &  0.9313 & 0.51& 5000  \\
$       $    &                         &                                                 &               &            &   &  &   &  \\
$10^{10}$   &      OB                &    $1024\times 1024$                   &   12            &    12        &  104.49 & 4344.77 &  0.5045& 0.89  & 800  \\
$10^{10}$   &      0                   &    $1024\times 1024$                   &    11           &   12         &  106.01 & 4481.58 &  0.5261 & 0.44 & 800  \\
$10^{10}$   &      0.1                   &    $1024\times 1024$                   &   12            &   13         &  99.09 & 4010.30 &  0.5290 & 0.88 & 800  \\
$10^{10}$   &      0.2                  &    $1024\times 1024$                   &    12           &   14         &  91.61 & 3504.57 &  0.5414 & 0.97 & 800  \\
$10^{10}$   &      0.3                   &    $1024\times 1024$                   &    13           &   16         & 82.50 & 2931.49 &  0.5598& 0.22 & 800  \\
$10^{10}$   &      0.4                   &    $1024\times 1024$                   &     14          &   20         &  69.35 & 2466.07 &  0.5951 & 0.23 & 800  \\
$10^{10}$   &      0.5                  &    $1024\times 1024$                   &   16            &   27         &  55.20 & 1971.33 &  0.6426 & 0.09& 800  \\
$10^{10}$   &      0.6                   &    $1024\times 1024$                   &   17            &   40         &  40.17 & 1541.42 &  0.7078 & 0.17 & 800  \\
$10^{10}$   &      0.7                   &    $1024\times 1024$                   &   20            &    66        & 25.72 & 1112.56 &  0.7810 & 0.67& 1600  \\
$10^{10}$   &      0.8                   &    $512\times 512$                   &      12         &   60         &  13.82 & 691.90 &  0.8560 & 0.22 & 3200  \\
$10^{10}$   &      0.9                   &    $512\times 512$                   &     17          &   138         &  4.77 & 286.17 &  0.9281& 0.42 & 2400  \\
$       $    &                         &                                                 &               &            &   &  &   & & \\
\end{tabular}
\end{center}
\caption{2D simulation parameters. The columns from left to right indicate Rayleigh number $Ra$, density inversion parameter $\theta_m$(OB cases are also included for comparison), grid resolutions, number of grid points in the bottom ($N_b$) and top ($N_t$) thermal boundary layers, Nusselt number at the bottom plate $Nu_b$, Reynolds number $Re$, centre temperature $\theta_c$, the difference between Nusselt number at the bottom and top plates ($\Delta Nu=|Nu_{b}-Nu_{t}|/Nu_{b}) $, and the averaging time for the simulations $t_{avg}$. Note that not all the 2D simulations in this work are listed in this table.}\label{tab2d}
\end{table}

\begin{table}
\tabcolsep 6pt
\renewcommand{\arraystretch}{1.}
\label{table1}
\begin{center}
\begin{tabular}{ccccccccccc}
$Ra$      &  $\Gamma_{3}$   & $\theta_m$   & $Nx\times Ny \times N_z$    &  $N_b$   & $N_t$  & $Nu_b$ &  $Re$ &  $\theta_c$  & 100$\Delta Nu$ &$t_{avg}$ \\
$10^8$   &   1/4   &OB                &    $64\times256\times 256$  &    9          &     9     &  31.64 & 135.63&   0.4987 &0.31 & 1000  \\
$10^8$   &   1/4   &0                   &    $64\times256\times 256$   &    9          &     9       &  31.86 & 139.23&  0.5229 &0.10 & 1000  \\
$10^8$   &   1/4   &0.1                   &    $64\times256\times 256$   & 9            & 10            &  29.94 & 124.77&  0.5275 &0.40 & 1000  \\
$10^8$   &   1/4   &0.2                   &    $64\times256\times 256$   & 10             &  11          &  27.75 & 109.95&  0.5345 &0.07 & 1000  \\
$10^8$   &   1/4   &0.3                   &    $64\times256\times 256$ &   10            &   12       &  25.08 & 95.16&   0.5497& 0.10 & 1000  \\
$10^8$   &   1/4   &0.4                   &    $64\times256\times 256$   & 11             &  15          &  21.02 & 77.90&  0.5814 &0.17 & 1000  \\
$10^8$   &   1/4   &0.5                  &    $64\times256\times 256$   &    12            &   21       &  16.13 & 63.14&   0.6458&  0.19& 1000  \\
$10^8$   &   1/4   &0.6                   &    $64\times256\times 256$   &    14          &   31         &  11.36 & 50.03&  0.5202 &0.36 & 1000  \\
$10^8$   &   1/4   &0.7                   &    $64\times256\times 256$  &     15            &   46         &  7.34 & 36.55&   0.7949&0.27  & 2000  \\
$       $    &     &                  &                                                 &               &            &   &  &   &&  \\
$10^8$   &   1/2   &OB                &    $128\times256\times 256$  &     9          &    9       &  31.21 & 165.85&    0.5036 &0.01 & 1000  \\
$10^8$   &   1/2   &0                   &    $128\times256\times 256$   &     9          &   10        &  31.75 & 170.99&  0.5277 &0.23 & 1000  \\
$10^8$   &   1/2   &0.1                   &    $128\times256\times 256$   &    9           &  10         &  29.77 & 154.09&  0.5285 &0.50 & 1000  \\
$10^8$   &   1/2   &0.2                   &    $128\times256\times 256$   &    9           & 11          &  27.53 & 136.12&  0.5444 &0.09 & 1000  \\
$10^8$   &   1/2   &0.3                   &    $128\times256\times 256$ &    10           &  13        &  24.89 & 116.55&   0.5576& 0.26 & 1000  \\
$10^8$   &   1/2   &0.4                   &    $128\times256\times 256$   &   11            &  15         &  21.38 & 95.72&  0.5939 &0.44 & 1000  \\
$10^8$   &   1/2   &0.5                  &    $128\times256\times 256$   &  12              &  20         &  16.36 & 76.28&   0.6373&  0.07& 1000  \\
$10^8$   &   1/2   &0.6                   &    $128\times256\times 256$   &   14            &   30        &  11.41 & 58.62&  0.7180 &0.55 & 1000  \\
$10^8$   &   1/2   &0.7                   &    $128\times256\times 256$  &   16                &  44          &  7.51 & 44.66&   0.7817&0.99  & 1000  \\
$       $    &     &                  &                                                 &               &            &   &  &   &&  \\
$10^8$   &   1   &OB                &    $256\times 256\times 256$  &    9         &     9       &  31.85 & 197.17&   0.4985 &0.10 & 1000  \\
$10^8$   &   1   &0                   &    $256\times 256\times 256$   &    9           &    9        &  32.00 & 200.52&  0.5250 &0.12 & 1000  \\
$10^8$   &   1   &0.1                   &    $256\times 256\times 256$   &   9            & 10           &  30.19 & 181.41&  0.5332 &0.24 & 1000  \\
$10^8$   &   1   &0.2                   &    $256\times 256\times 256$   &   9            &  11          &  27.95 & 162.68&  0.5424 &0.55 & 1000  \\
$10^8$   &   1   &0.3                   &    $256\times 256\times 256$ &    10            &  12       &  25.46 & 137.66&   0.5567& 0.99 & 1000  \\
$10^8$   &   1   &0.4                   &    $256\times 256\times 256$   &   11            &   15         & 21.80 &116.22&  0.5956 &0.60 & 1000  \\
$10^8$   &   1   &0.5                  &    $256\times 256\times 256$   &     12            &   20        &  16.82 & 91.53&   0.6463& 0.75 & 1000  \\
$10^8$   &   1   &0.6                   &    $256\times 256\times 256$   &    13           &   30         &  11.90 & 71.11&  0.7167 &0.32 & 1000  \\
$10^8$   &   1   &0.7                   &    $256\times 256\times 256$  &     15              &  44         &  7.71 & 52.23&   0.7895&0.91  & 1000  \\
$10^8$   &   1   &0.8                   &    $256\times 256\times 256$   &   9            &    36        &  4.18 & 34.66&  0.8604 &0.38 & 1000  \\
$       $    &     &                  &                                                 &               &            &   &  &   &&  \\
$10^9$   &  1/2   &OB                &    $256\times 512\times 512$  &   10          &   10         &  62.84 & 595.02&   0.5011 &0.28 & 800  \\
$10^9$   &  1/2   &0                   &    $256\times 512\times 512$   &   9           &   10         &  63.41 & 610.08&  0.5286 &0.44 & 800  \\
$10^9$   &   1/2   &0.1                   &    $256\times 512\times 512$   & 9              &  11          &  59.49 & 551.32&  0.5333 &0.04 & 800  \\
$10^9$   &   1/2   &0.2                   &    $256\times 512\times 512$   &  10             &  12          &  54.81 & 482.42&  0.5428 &0.70 & 800  \\
$10^9$   &   1/2   &0.3                   &    $256\times 512\times 512$ &   10             &   13      &  49.54 & 421.28&   0.5622& 0.18 & 800  \\
$10^9$   &   1/2   &0.4                   &    $256\times 512\times 512$   & 11              &   16         & 42.52 &351.36&  0.5962 &0.08 & 800  \\
$10^9$   &   1/2   &0.5                  &    $256\times 512\times 512$   &   12              &  22        &  33.28 &282.92&   0.6475& 0.10 & 800  \\
$10^9$   &   1/2   &0.6                   &    $256\times 512\times 512$   &  14             &    33        &  23.49 & 220.98&  0.7149 &0.06 & 800  \\
$10^9$   &   1/2   &0.7                   &    $256\times 512\times 512$  &   16                &  53         & 15.13 & 155.12&   0.7884&0.48  & 800  \\
$10^9$   &   1/2   &0.8                   &    $128\times 256\times 256$   &  10             &  46          &  7.97 & 100.01&  0.8585 &0.44 & 1000  \\
$10^9$   &   1/2   &0.9                   &    $128\times 256\times 256$   &   14          &    100        &  2.80 & 42.69&  0.9312 &0.17 & 1000  \\
$       $    &     &                  &                                                 &               &            &   &  &   &&  \\
\end{tabular}
\end{center}
\caption{3D simulation parameters. The columns from left to right indicate Rayleigh number $Ra$, density inversion parameter $\theta_m$(OB cases are also included for comparison), grid resolutions, number of grid points in the bottom ($N_b$) and top ($N_t$) thermal boundary layers, Nusselt number at the bottom plate $Nu_b$, Reynolds number $Re$, centre temperature $\theta_c$, the difference between Nusselt number at the bottom and top plates ($\Delta Nu=|Nu_{b}-Nu_{t}|/Nu_{b}) $, and the averaging time for the simulations $t_{avg}$.  }\label{sim3d}
\end{table}

\section*{Acknowledgements}
We wish to thank Kai Leong Chong for helpful discussions. This work is supported by National Natural Science Foundation of China under Grants (11572314, 11772323, 11621202 and 11825204), and the Fundamental Research Funds for the Central Universities.
\bibliographystyle{jfm}
\bibliography{penetrate}

\begin{thebibliography}{49}
\expandafter\ifx\csname natexlab\endcsname\relax\def\natexlab#1{#1}\fi
\def\au#1{#1} \def\ed#1{#1} \def\yr#1{#1}\def\at#1{#1}\def\jt#1{\textit{#1}}
  \def\bt#1{#1}\def\bvol#1{\textbf{#1}} \def\vol#1{#1} \def\pg#1{#1}
  \def\publ#1{#1}\def\arxiv#1{#1}\def\org#1{#1}\def\st#1{\textit{#1}}

\bibitem[Ahlers {\em et~al.\/}(2007)Ahlers, Araujo, Funfschilling, Grossmann \&
  Lohse]{ahlers2007non}
{\sc \au{Ahlers, G.}, \au{Araujo, F.~F.}, \au{Funfschilling, D.},
  \au{Grossmann, S.} \& \au{Lohse, D.}} \yr{2007}  \at{{Non-Oberbeck-Boussinesq
  effects in gaseous Rayleigh-B{\'e}nard convection}}.  \jt{Phys. Rev. Lett.}
  \bvol{98}~(5),  \pg{054501}.

\bibitem[Ahlers {\em et~al.\/}(2006)Ahlers, Brown, Araujo, Funfschilling,
  Grossmann \& Lohse]{ahlers2006non}
{\sc \au{Ahlers, G.}, \au{Brown, E.}, \au{Araujo, F.~F.}, \au{Funfschilling,
  D.}, \au{Grossmann, S.} \& \au{Lohse, D.}} \yr{2006}
  \at{{Non-Oberbeck--Boussinesq effects in strongly turbulent
  Rayleigh--B{\'e}nard convection}}.  \jt{J. Fluid Mech.}  \bvol{569},
  \pg{409--445}.

\bibitem[Ahlers {\em et~al.\/}(2009)Ahlers, Grossmann \& Lohse]{ahlers2009heat}
{\sc \au{Ahlers, G.}, \au{Grossmann, S.} \& \au{Lohse, D.}} \yr{2009}
  \at{{Heat transfer and large scale dynamics in turbulent Rayleigh-B{\'e}nard
  convection}}.  \jt{Rev. Mod. Phys.}  \bvol{81},  \pg{503--537}.

\bibitem[Antar(1987)]{antar1987penetrative}
{\sc \au{Antar, B.~N.}} \yr{1987}  \at{Penetrative double-diffusive
  convection}.  \jt{Phys. Fluids}  \bvol{30}~(2),  \pg{322--330}.

\bibitem[Buffett(2014)]{buffett2014geomagnetic}
{\sc \au{Buffett, B.}} \yr{2014}  \at{{Geomagnetic fluctuations reveal stable
  stratification at the top of the Earth's core}}.  \jt{Nature}
  \bvol{507}~(7493),  \pg{484}.

\bibitem[Chill{\`a} \& Schumacher(2012)]{chilla2012new}
{\sc \au{Chill{\`a}, F} \& \au{Schumacher, J}} \yr{2012}  \at{{New perspectives
  in turbulent Rayleigh-B{\'e}nard convection}}.  \jt{Eur. Phys. J. E}
  \bvol{35}~(7),  \pg{58}.

\bibitem[Chong {\em et~al.\/}(2015)Chong, Huang, Kaczorowski \&
  Xia]{chong2015condensation}
{\sc \au{Chong, K.~L.}, \au{Huang, S.-D.}, \au{Kaczorowski, M.} \& \au{Xia,
  K.-Q.}} \yr{2015}  \at{Condensation of coherent structures in turbulent
  flows}.  \jt{Phys. Rev. Lett.}  \bvol{115}~(26),  \pg{264503}.

\bibitem[Chong {\em et~al.\/}(2018)Chong, Wagner, Kaczorowski, Shishkina \&
  Xia]{chong2018effect}
{\sc \au{Chong, K.~L.}, \au{Wagner, S.}, \au{Kaczorowski, M.}, \au{Shishkina,
  O.} \& \au{Xia, K.-Q.}} \yr{2018}  \at{{Effect of Prandtl number on heat
  transport enhancement in Rayleigh-B{\'e}nard convection under geometrical
  confinement}}.  \jt{Phys. Rev. Fluid}  \bvol{3}~(1),  \pg{013501}.

\bibitem[Chong \& Xia(2016)]{chong2016exploring}
{\sc \au{Chong, K.~L.} \& \au{Xia, K.-Q.}} \yr{2016}  \at{{Exploring the
  severely confined regime in Rayleigh--B{\'e}nard convection}}.  \jt{J. Fluid
  Mech.}  \bvol{805}.

\bibitem[Chong {\em et~al.\/}(2017)Chong, Yang, Huang, Zhong, Stevens,
  Verzicco, Lohse \& Xia]{chong2017confined}
{\sc \au{Chong, K.~L.}, \au{Yang, Y.-T.}, \au{Huang, S.-D.}, \au{Zhong, J.-Q.},
  \au{Stevens, R. J. A.~M.}, \au{Verzicco, R.}, \au{Lohse, D.} \& \au{Xia,
  K.-Q.}} \yr{2017}  \at{{Confined Rayleigh-B{\'e}nard, Rotating
  Rayleigh-B{\'e}nard, and Double Diffusive Convection: A unifying view on
  turbulent transport enhancement through coherent structure manipulation}}.
  \jt{Phys. Rev. Lett.}  \bvol{119}~(6),  \pg{064501}.

\bibitem[Couston {\em et~al.\/}(2018)Couston, Lecoanet, Favier \&
  Le~Bars]{couston2018order}
{\sc \au{Couston, L.-A.}, \au{Lecoanet, D.}, \au{Favier, B.} \& \au{Le~Bars,
  M.}} \yr{2018}  \at{Order out of chaos: Slowly reversing mean flows emerge
  from turbulently generated internal waves}.  \jt{Phys. Rev. Lett.}
  \bvol{120}~(24),  \pg{244505}.

\bibitem[Dintrans {\em et~al.\/}(2005)Dintrans, Brandenburg, Nordlund \&
  Stein]{dintrans2005spectrum}
{\sc \au{Dintrans, B.}, \au{Brandenburg, A.}, \au{Nordlund, {\AA}.} \&
  \au{Stein, R.~F.}} \yr{2005}  \at{Spectrum and amplitudes of internal gravity
  waves excited by penetrative convection in solar-type stars}.  \jt{Astron.
  Astrophys}  \bvol{438}~(1),  \pg{365--376}.

\bibitem[Gebhart \& Mollendorf(1977)]{gebhart1977new}
{\sc \au{Gebhart, B.} \& \au{Mollendorf, J.~C.}} \yr{1977}  \at{A new density
  relation for pure and saline water}.  \jt{Deep-Sea Res.}  \bvol{24}~(9),
  \pg{831--848}.

\bibitem[Goluskin \& van~der Poel(2016)]{goluskin2016penetrative}
{\sc \au{Goluskin, D.} \& \au{van~der Poel, E.~P.}} \yr{2016}  \at{Penetrative
  internally heated convection in two and three dimensions}.  \jt{J. Fluid
  Mech.}  \bvol{791}.

\bibitem[Horn \& Shishkina(2014)]{horn2014rotating}
{\sc \au{Horn, S.} \& \au{Shishkina, O.}} \yr{2014}  \at{{Rotating
  non-Oberbeck--Boussinesq Rayleigh--B{\'e}nard convection in water}}.
  \jt{Phys. Fluids}  \bvol{26}~(5),  \pg{055111}.

\bibitem[Horn {\em et~al.\/}(2013)Horn, Shishkina \& Wagner]{horn2013non}
{\sc \au{Horn, S.}, \au{Shishkina, O.} \& \au{Wagner, C.}} \yr{2013}  \at{{On
  non-Oberbeck--Boussinesq effects in three-dimensional Rayleigh--B{\'e}nard
  convection in glycerol}}.  \jt{J. Fluid Mech.}  \bvol{724},  \pg{175--202}.

\bibitem[Hu {\em et~al.\/}(2015)Hu, Li \& Wu]{hu2015rayleigh}
{\sc \au{Hu, Y.-P.}, \au{Li, Y.-R.} \& \au{Wu, C.-M.}} \yr{2015}
  \at{{Rayleigh-B{\'e}nard convection of cold water near its density maximum in
  a cubical cavity}}.  \jt{Phys. Fluids}  \bvol{27}~(3),  \pg{034102}.

\bibitem[Huang {\em et~al.\/}(2013)Huang, Kaczorowski, Ni \&
  Xia]{huang2013confinement}
{\sc \au{Huang, S.-D.}, \au{Kaczorowski, M.}, \au{Ni, R.} \& \au{Xia, K.-Q.}}
  \yr{2013}  \at{Confinement-induced heat-transport enhancement in turbulent
  thermal convection}.  \jt{Phys. Rev. Lett.}  \bvol{111}~(10),  \pg{104501}.

\bibitem[Huang \& Zhou(2013)]{huang2013counter}
{\sc \au{Huang, Y.-X.} \& \au{Zhou, Q.}} \yr{2013}  \at{{Counter-gradient heat
  transport in two-dimensional turbulent Rayleigh--B{\'e}nard convection}}.
  \jt{J. Fluid Mech.}  \bvol{737}.

\bibitem[Johnston \& Doering(2009)]{johnston2009comparison}
{\sc \au{Johnston, H.} \& \au{Doering, C.~R.}} \yr{2009}  \at{Comparison of
  turbulent thermal convection between conditions of constant temperature and
  constant flux}.  \jt{Phys. Rev. Lett.}  \bvol{102}~(6),  \pg{064501}.

\bibitem[Kong {\em et~al.\/}(2018)Kong, Zhang, Schubert \&
  Anderson]{kong2018origin}
{\sc \au{Kong, Dali}, \au{Zhang, Keke}, \au{Schubert, Gerald} \& \au{Anderson,
  John~D}} \yr{2018}  \at{Origin of jupiter’s cloud-level zonal winds remains
  a puzzle even after juno}.  \jt{Proc. N. Acad. Sci.}  \bvol{115}~(34),
  \pg{8499--8504}.

\bibitem[Large \& Andereck(2014)]{large2014penetrative}
{\sc \au{Large, E} \& \au{Andereck, CD}} \yr{2014}  \at{{Penetrative
  Rayleigh-B{\'e}nard convection in water near its maximum density point}}.
  \jt{Phys. Fluids}  \bvol{26}~(9),  \pg{094101}.

\bibitem[Lecoanet {\em et~al.\/}(2015)Lecoanet, Le~Bars, Burns, Vasil, Brown,
  Quataert \& Oishi]{lecoanet2015numerical}
{\sc \au{Lecoanet, D.}, \au{Le~Bars, M.}, \au{Burns, K.~J.}, \au{Vasil, G.~M.},
  \au{Brown, B.~P.}, \au{Quataert, E.} \& \au{Oishi, J.~S.}} \yr{2015}
  \at{Numerical simulations of internal wave generation by convection in
  water}.  \jt{Phys. Rev. E}  \bvol{91}~(6),  \pg{063016}.

\bibitem[Leighton(1963)]{leighton1963solar}
{\sc \au{Leighton, R.~B.}} \yr{1963}  \at{The solar granulation}.  \jt{Annu.
  Rev. Astron. and Astrophys}  \bvol{1}~(1),  \pg{19--40}.

\bibitem[Liu {\em et~al.\/}(2018)Liu, Xia, Yan, Wan \& Sun]{liu2018linear}
{\sc \au{Liu, S.}, \au{Xia, S.-N.}, \au{Yan, R.}, \au{Wan, Z.-H.} \& \au{Sun,
  D.-J.}} \yr{2018}  \at{{Linear and weakly nonlinear analysis of
  Rayleigh--B{\'e}nard convection of perfect gas with non-Oberbeck--Boussinesq
  effects}}.  \jt{J. Fluid Mech.}  \bvol{845},  \pg{141--169}.

\bibitem[Lohse \& Xia(2010)]{lohse2010small}
{\sc \au{Lohse, D.} \& \au{Xia, K.-Q.}} \yr{2010}  \at{{Small-scale properties
  of turbulent Rayleigh-B{\'e}nard convection}}.  \jt{Annu. Rev. Fluid Mech}
  \bvol{42},  \pg{335--364}.

\bibitem[Moore \& Weiss(1973)]{moore1973nonlinear}
{\sc \au{Moore, DR} \& \au{Weiss, NO}} \yr{1973}  \at{Nonlinear penetrative
  convection}.  \jt{J. Fluid Mech.}  \bvol{61}~(3),  \pg{553--581}.

\bibitem[Musman(1968)]{musman1968penetrative}
{\sc \au{Musman, S.}} \yr{1968}  \at{Penetrative convection}.  \jt{J. Fluid
  Mech.}  \bvol{31}~(2),  \pg{343--360}.

\bibitem[Ostilla-M{\'o}nico {\em et~al.\/}(2014)Ostilla-M{\'o}nico, van~der
  Poel, Verzicco, Grossmann \& Lohse]{ostilla2014exploring}
{\sc \au{Ostilla-M{\'o}nico, R.}, \au{van~der Poel, E.~P.}, \au{Verzicco, R.},
  \au{Grossmann, S.} \& \au{Lohse, D.}} \yr{2014}  \at{{Exploring the phase
  diagram of fully turbulent Taylor--Couette flow}}.  \jt{J. Fluid Mech.}
  \bvol{761},  \pg{1--26}.

\bibitem[Ostilla-M{\'o}nico {\em et~al.\/}(2013)Ostilla-M{\'o}nico, Stevens,
  Grossmann, Verzicco \& Lohse]{ostilla2013optimal}
{\sc \au{Ostilla-M{\'o}nico, R.}, \au{Stevens, R. J. A.~M}, \au{Grossmann, S.},
  \au{Verzicco, R.} \& \au{Lohse, D.}} \yr{2013}  \at{{Optimal Taylor--Couette
  flow: direct numerical simulations}}.  \jt{J. Fluid Mech.}  \bvol{719},
  \pg{14--46}.

\bibitem[Sameen {\em et~al.\/}(2008)Sameen, Verzicco \&
  Sreenivasan]{sameen2008non}
{\sc \au{Sameen, A}, \au{Verzicco, R} \& \au{Sreenivasan, KR}} \yr{2008}
  \at{Non-boussinesq convection at moderate rayleigh numbers in low temperature
  gaseous helium}.  \jt{Phys. Scr.}  \bvol{2008}~(T132),  \pg{014053}.

\bibitem[Sameen {\em et~al.\/}(2009)Sameen, Verzicco \&
  Sreenivasan]{sameen2009specific}
{\sc \au{Sameen, A}, \au{Verzicco, R} \& \au{Sreenivasan, KR}} \yr{2009}
  \at{Specific roles of fluid properties in non-boussinesq thermal convection
  at the rayleigh number of 2$\times 10^8$}.  \jt{Eur. Phys. Lett.}
  \bvol{86}~(1),  \pg{14006}.

\bibitem[Shishkina {\em et~al.\/}(2010)Shishkina, Stevens, Grossmann \&
  Lohse]{shishkina2010boundary}
{\sc \au{Shishkina, O.}, \au{Stevens, R.J.A.M.}, \au{Grossmann, S.} \&
  \au{Lohse, D.}} \yr{2010}  \at{Boundary layer structure in turbulent thermal
  convection and its consequences for the required numerical resolution}.
  \jt{New J. Phys.}  \bvol{12}~(7),  \pg{075022}.

\bibitem[Sugiyama {\em et~al.\/}(2009)Sugiyama, Calzavarini, Grossmann \&
  Lohse]{sugiyama2009flow}
{\sc \au{Sugiyama, K.}, \au{Calzavarini, E.}, \au{Grossmann, S.} \& \au{Lohse,
  D.}} \yr{2009}  \at{{Flow organization in two-dimensional
  non-Oberbeck--Boussinesq Rayleigh--B{\'e}nard convection in water}}.  \jt{J.
  Fluid Mech.}  \bvol{637},  \pg{105--135}.

\bibitem[Sugiyama {\em et~al.\/}(2010)Sugiyama, Ni, Stevens, Chan, Zhou, Xi,
  Sun, Grossmann, Xia \& Lohse]{sugiyama2010flow}
{\sc \au{Sugiyama, K.}, \au{Ni, R.}, \au{Stevens, R. J. A.~M.}, \au{Chan,
  T.~S.}, \au{Zhou, S.-Q.}, \au{Xi, H.-D.}, \au{Sun, C.}, \au{Grossmann, S.},
  \au{Xia, K.-Q.} \& \au{Lohse, D.}} \yr{2010}  \at{Flow reversals in thermally
  driven turbulence}.  \jt{Phys. Rev. Lett.}  \bvol{105}~(3),  \pg{034503}.

\bibitem[Toppaladoddi \& Wettlaufer(2018)]{toppaladoddi2018penetrative}
{\sc \au{Toppaladoddi, S.} \& \au{Wettlaufer, J.~S.}} \yr{2018}
  \at{{Penetrative convection at high Rayleigh numbers}}.  \jt{Phys. Rev.
  Fluid}  \bvol{3}~(4),  \pg{043501}.

\bibitem[Van~Gils {\em et~al.\/}(2011)Van~Gils, Huisman, Bruggert, Sun \&
  Lohse]{van2011torque}
{\sc \au{Van~Gils, D. P.~M.}, \au{Huisman, S.~G.}, \au{Bruggert, G.-W.},
  \au{Sun, C.} \& \au{Lohse, D.}} \yr{2011}  \at{{Torque scaling in turbulent
  Taylor-Couette flow with co-and counterrotating cylinders}}.  \jt{Phys. Rev.
  Lett.}  \bvol{106}~(2),  \pg{024502}.

\bibitem[Veronis(1963)]{veronis1963penetrative}
{\sc \au{Veronis, George}} \yr{1963}  \at{Penetrative convection.}
  \jt{Astrophys. J.}  \bvol{137},  \pg{641}.

\bibitem[Wang {\em et~al.\/}(2018{\natexlab{{\em a\/}}})Wang, Wan, Yan \&
  Sun]{wang2018multiple}
{\sc \au{Wang, Q.}, \au{Wan, Z.-H.}, \au{Yan, R.} \& \au{Sun, D.-J.}}
  \yr{2018{\natexlab{{\em a\/}}}}  \at{Multiple states and heat transfer in
  two-dimensional tilted convection with large aspect ratios}.  \jt{Phys. Rev.
  Fluid}  \bvol{3}~(11),  \pg{113503}.

\bibitem[Wang {\em et~al.\/}(2019{\natexlab{{\em a\/}}})Wang, Wan, Yan \&
  Sun]{wang2019flow}
{\sc \au{Wang, Q.}, \au{Wan, Z.-H.}, \au{Yan, R.} \& \au{Sun, D.-J.}}
  \yr{2019{\natexlab{{\em a\/}}}}  \at{Flow organization and heat transfer in
  two-dimensional tilted convection with aspect ratio 0.5}.  \jt{Phys. Fluids}
  \bvol{31}~(2),  \pg{025102}.

\bibitem[Wang {\em et~al.\/}(2018{\natexlab{{\em b\/}}})Wang, Xia, Wang, Sun,
  Zhou \& Wan]{wang2018flow}
{\sc \au{Wang, Q.}, \au{Xia, S.-N.}, \au{Wang, B.-F.}, \au{Sun, D.-J.},
  \au{Zhou, Q.} \& \au{Wan, Z.-H.}} \yr{2018{\natexlab{{\em b\/}}}}  \at{Flow
  reversals in two-dimensional thermal convection in tilted cells}.  \jt{J.
  Fluid Mech.}  \bvol{849},  \pg{355--372}.

\bibitem[Wang {\em et~al.\/}(2019{\natexlab{{\em b\/}}})Wang, Xia, Yan, Sun \&
  Wan]{wang2019non}
{\sc \au{Wang, Q.}, \au{Xia, S.-N.}, \au{Yan, R.}, \au{Sun, D.-J.} \& \au{Wan,
  Z.-H.}} \yr{2019{\natexlab{{\em b\/}}}}  \at{Non-oberbeck-boussinesq effects
  due to large temperature differences in a differentially heated square cavity
  filled with air}.  \jt{Int. J. Heat Mass Transf.}  \bvol{128},
  \pg{479--491}.

\bibitem[Wang {\em et~al.\/}(2017)Wang, Xu, Xia, Wan \& Sun]{wang2017thermal}
{\sc \au{Wang, Q.}, \au{Xu, B.-L.}, \au{Xia, S.-N.}, \au{Wan, Z.-H.} \&
  \au{Sun, D.-J.}} \yr{2017}  \at{Thermal convection in a tilted rectangular
  cell with aspect ratio 0.5}.  \jt{Chin. Phys. Lett.}  \bvol{34}~(10),
  \pg{104401}.

\bibitem[Weiss {\em et~al.\/}(2018)Weiss, He, Ahlers, Bodenschatz \&
  Shishkina]{weiss2018bulk}
{\sc \au{Weiss, S.}, \au{He, X.}, \au{Ahlers, G.}, \au{Bodenschatz, E.} \&
  \au{Shishkina, O.}} \yr{2018}  \at{{Bulk temperature and heat transport in
  turbulent Rayleigh--B{\'e}nard convection of fluids with
  temperature-dependent properties}}.  \jt{J. Fluid Mech.}  \bvol{851},
  \pg{374--390}.

\bibitem[Xia {\em et~al.\/}(2016)Xia, Wan, Liu, Wang \& Sun]{xia2016flow}
{\sc \au{Xia, S.-N.}, \au{Wan, Z.-H.}, \au{Liu, S.}, \au{Wang, Q.} \& \au{Sun,
  D.-J.}} \yr{2016}  \at{{Flow reversals in Rayleigh--B{\'e}nard convection
  with non-Oberbeck--Boussinesq effects}}.  \jt{J. Fluid Mech.}  \bvol{798},
  \pg{628--642}.

\bibitem[Zhang {\em et~al.\/}(1997)Zhang, Childress \& Libchaber]{zhang1997non}
{\sc \au{Zhang, J.}, \au{Childress, S.} \& \au{Libchaber, A.}} \yr{1997}
  \at{Non-boussinesq effect: Thermal convection with broken symmetry}.
  \jt{Phys. Fluids}  \bvol{9}~(4),  \pg{1034--1042}.

\bibitem[Zhang \& Schubert(1996)]{zhang1996penetrative}
{\sc \au{Zhang, K.K.} \& \au{Schubert, G.}} \yr{1996}  \at{Penetrative
  convection and zonal flow on jupiter}.  \jt{Science}  \bvol{273}~(5277),
  \pg{941--943}.

\bibitem[Zhang \& Schubert(2000)]{zhang2000teleconvection}
{\sc \au{Zhang, K.K.} \& \au{Schubert, G.}} \yr{2000}  \at{Teleconvection:
  remotely driven thermal convection in rotating stratified spherical layers}.
  \jt{Science}  \bvol{290}~(5498),  \pg{1944--1947}.

\bibitem[Zhang {\em et~al.\/}(2017)Zhang, Zhou \& Sun]{zhang2017statistics}
{\sc \au{Zhang, Y.}, \au{Zhou, Q.} \& \au{Sun, C.}} \yr{2017}  \at{{Statistics
  of kinetic and thermal energy dissipation rates in two-dimensional turbulent
  Rayleigh--B{\'e}nard convection}}.  \jt{J. Fluid Mech.}  \bvol{814},
  \pg{165--184}.

\end{thebibliography}

\end{document}